\documentclass[preprint,12pt,authoryear]{elsarticle}
 \usepackage{graphicx}
\usepackage{amsmath}
\usepackage{amsfonts}
\usepackage{amssymb}
\usepackage{amsthm,mathrsfs}
\usepackage{color}
\usepackage{physics}
\usepackage{xr-hyper}
 \usepackage{hyperref}

 \newcommand{\be}{\begin{equation}}
 \newcommand{\ee}{\end{equation}}
 \newcommand{\ba}{\begin{eqnarray}}
 \newcommand{\ea}{\end{eqnarray}}
 \newcommand{\efbold}{\mbox{{\boldmath $\vec f$}}}
 
 \newcommand{\erbold}{\mbox{{\boldmath $\vec r$}}}
 \newcommand{\Phibold}{\mbox{{\boldmath $\vec \Phi$}}}

 \newcommand{\ddoterbold}{\ddot{\textbf {\mbox{\boldmath $\vec {\boldmath r}$}}}}
 \newcommand{\doterbold}{\dot{\textbf {\mbox{\boldmath $\vec{\boldmath r}$}} }}

 \newcommand{\km}{\textrm{km}}
 \newcommand{\rad}{\textrm{rad}}

 \newcommand{\p}{\partial}
\newcommand{\ds}{\displaystyle}
\newcommand{\cc}{\textrm{c}}
\newcommand{\s}{\textrm{s}}
\newcommand{\ve}{\varepsilon}
 
\renewcommand{\vec}[1]{{\boldsymbol{\mathrm{#1}}}}

\journal{Advances in Space Research}

\begin{document}
\begin{frontmatter}
\pdfbookmark[1]{Front Matter}{frontmatter}

 \title{
    {\small{~~~~~~\qquad \qquad\qquad Published in \,{\it{Advances in Space Research}},\, Vol.  ,\, pp.  \,(2021) }}\\
    ~\\
 {Analysis of the PPN Two-Body Problem}\\ {Using Non-Osculating Orbital Elements}}

 \author{ {\large Pini Gurfil}\\
 Faculty of Aerospace Engineering, Technion, Haifa 3200003 Israel\\
 pgurfil@technion.ac.il
 }

\author{~\\ ~\\ {\large Michael Efroimsky}\\ 
US Naval Observatory, Washington DC 20392 USA\\
{michael.efroimsky@gmail.com}
}

%
%


%
 \begin{abstract}
 The parameterised post-Newtonian (PPN) formalism is a weak-field and slow-motion approximation for both General Relativity (GR) and some of its viable generalisations. Within this formalism, the motion can be approached using various parameterisations, among which are the Lagrange-type and Gauss-type orbital equations. Often, these equations are developed under the premise of the Lagrange constraint. This constraint makes the evolving orbital elements parameterise instantaneous conics always tangent to the actual trajectory. Arbitrary mathematically, this choice of a constraint is convenient under perturbations dependent only on positions. However, under perturbations dependent also on velocities (like in the  relativistic celestial mechanics) the Lagrange constraint unnecessarily complicates solutions that can be simplified by relaxing the constraint and introducing a freedom in the orbit parameterisation, which is analogous to the gauge freedom in electrodynamics and gauge field theories. Geometrically, this freedom is the freedom of nonosculation, i.e., of the degree to which the instantaneous conics are permitted to be non-tangent to the actual orbit. Under the same perturbation, all solutions with different degree of nonosculation look mathematically different, though describe the same physical orbit. While non-intuitive, the modeling of an orbit with a sequence of nontangent instantaneous conics can at times simplify calculations. The  appropriately generalised (``gauge-generalised'') Lagrange-type equations, and their applications, appeared in the literature heretofore. We, in this paper, derive the gauge-generalised Gauss-type equations and apply them to the PPN two-body problem.
 Fixing the gauge freedom in three different ways (i.e., modeling an orbit with non-osculating elements of three different types) we find three parameterisations of the PPN two-body dynamics. These parameterisations render orbits with either a fixed non-osculating semimajor axis, or a fixed non-osculating eccentricity, or a fixed non-osculating argument of periastron. We also develop a transformation from non-osculating to classical osculating orbital elements, and illustrate the new solutions using numerical simulations.
\end{abstract}


\end{frontmatter}

 \section{
 Introduction
 }

The oral tradition attributes to John Wheeler the following formulation of the essence of general relativity (GR):
  \footnote{~{For recent critical overviews of GR see, e.g., \citep{Iorio:2015mga,universe2040023,universe2020011,universe5070173} and GR centennial jubilee volumes edited by \citet{+2015} and  \citet{ashtekar2005100}.
 }}
``Spacetime tells matter how to move; matter tells spacetime how to curve.''
 A possible footnote to this proverbial quote could be that in the zero-curvature limit matter keeps moving --- though not necessarily in a Newtonian manner, because motion can be fast (luminal, for photons). Hence the question: would a {\it{slightly}} curved spacetime tell matter how to deviate {\it{slightly}} from Newtonian trajectories? A short answer to this question is partially positive: the zero-curvature slow-motion limit of GR is Newtonian; and there exists an approximation of GR, known as the post-Newtonian (PN) formalism, intended to adjust {\it{some}} GR problems to the Newtonian framework {\citep{Will5938}}. But then, it turns out that this approximation is fraught with obstacles.

 {A natural attempt to fit GR into a purely Newtonian framework is to consider the $N$-body problem in an asymptotically flat spacetime
 covered with a single global coordinate map. This treatment bears a trace of Newton's absolute space and time. However, practical needs (like the relativistic treatment of tidal forces) require introduction of $N + 1$ coordinate patches: a comoving patch for each body, and a global patch \citep{kopejkin1988}.
 The treatment then implies writing down the relations linking each local coordinate system to the global one \citep{1987thyg.book..128D}.
 This development was pioneered by \citet{brumberg1989} and was later extended by \citet{damour1991} who included all multipole moments in the expansion of the gravitational field.
 }

 The second major difficulty of the PN approach is describing post-Newtonian motion of {\it{extended}} bodies. While in the Newtonian theory the motion of spherically-symmetric bodies is identical to that of point particles, this identity holds in PN only for the translational motion of nonrotating spherically symmetrical bodies. So, for modeling the translational motion of such bodies, a knowledge of their masses is enough~--- a so-called effacing principle \citep{kopeikin_vlasov_2006}.
  But, in the presence of rotation or for curved trajectories, the description requires knowledge not only of their masses but also of their multipole momenta { \citep{kopeikin1985,2014CQGra..31x5012P,Meichsner:2015zma,1988CeMec..42...81S,2015IJMPD..2450067I,universe5070165}}.

  Despite these challenges, the PN
 formalism is applicable to various physical settings and renders numerous valuable results. According to \citet{blanchet2014}, this formalism has three principal applications:

\begin{itemize}

 \item[1.] Modeling of the $N$-body solar system dynamics (i.e. of the motion of the planets' centres of mass) {incorporates the first PN approximation, that of the order $(v/c)^2\;$ \citep{2015IJMPD..2450067I}.}

  \item[2.] The description of the gravitational radiation-reaction force emerging in pulsar dynamics employs the 2.5PN, i.e.
    order-$(v/c)^5\,$ equations { \citep{1983SvAL....9..230G,damour1983,iorio2019,iorio2021}}. The model has been experimentally
    verified via observations of the secular acceleration of the orbital motion of the Hulse-Taylor binary pulsar PSR 1913+16
    \citep{taylor1992, taylor1993} and of the pulsar PSR J0737-3039 \citep{possenti}.

  \item[3.] The equations of motion and the expressions for the radiation field, written down to higher PN orders, show up
    in the analysis of gravitational waves emitted by inspiralling compact binaries, cosmic pairs comprising two black holes, or two neutron stars, or a black hole and a neutron star \citep{LIGO}, driven into coalescence by emission of gravitational waves { \citep{universe2030022,2019Natur.568..469M,Bailes:2021tot}}.

\end{itemize}

 Addressing the first of these three applications, for point masses, our paper deals with the first post-Newtonian (1PN) approximation, one taking care of terms up to $(v/c)^2$ inclusively.


  The modern PN theory is fit to approximate the GR along with some of its generalisations. This tool
  is called the {\it{Parameterised post-Newtonian}} formalism, or PPN  {\citep{2003gr.qc.....4014B,Will:2018bme,Asada:1997yfu}}.  Within this parameterisation, the relativistic force can be expressed through a variation of the Lagrangian. One such force model was developed by \citet{chazy} and \citet{brumberg}, another by \citet{dd}.

 Our goal in this paper is to find solutions to the PPN two-body problem, in non-osculating elements (the bodies assumed spherically symmetric and non-rotating). In one such solution, nonosculation is picked up in such a way that the non-osculating semimajor axis stays fixed. In another solution, nonosculation is such that the non-osculating eccentricity stays constant. In the third solution, nonosculation is chosen in such a manner that the non-osculating argument of periastron is fixed. Under the so-arranged parameterisations, orbit integration gets simplified.

 {Be mindful that in our parameterisations it is the total rates (not just their secular parts) that are nullified.}

 \section{Methods}

  When a perturbing force is introduced into the Newtonian two-body setting, {the classical variation-of-parameters method suggest to recast the
  orbital elements as functions of time}. The resulting differential equations, describing the temporal change of the Keplerian orbital elements for a conservative, position-only dependent perturbing potential, are known as the \emph{orbital equations in the form of Lagrange}. On the other hand, the orbital equations in the form of Gauss, also known as the \emph{Gauss variational equations} or GVE, model the time evolution of the orbital elements due to an arbitrary perturbing force.

  For details of the method, we refer the reader to Appendix A,  while a more comprehensive explanation can be found in Efroimsky (\citeyear{efroim1}) and Efroimsky (\citeyear{efroim2}).
  In short, the method works as follows.  The perturbed vectorial equation of motion has three projections, which are three scalar second-order differential equations for the three Cartesian coordinates. In the Cauchy form, they render six first-order equations for the three coordinates $x^{\alpha}$ and three velocities $v^{\alpha}$. These are the three projections of the equation of motion, reading as $dv^{\alpha}/dt = \varphi(C_1\,,\,...\,,\,C_6\,,\,
  \dot{C}_1\,,\,...\,,\,\dot{C}_6)\,$, and the three obvious equations $dx^{\alpha}/dt = v^{\alpha}\,$. Be mindful that the insertion of $\,x^{\alpha} = f^{\alpha}(C_1(t)\,,\,...\,\,C_6(t)\,)\,$ in the equation of motion renders three (for
  $\,\alpha = 1,\,2,\,3\,$) scalar equations for the twelve variables $\,C_i(t)\,$ and $\,\dot{C}_i(t)\,$. (We get $\,\dot{C}_i\,$ because the ``constants'' are time dependent). If we now treat $\,\dot{C}_i\,$ as independent variables, and amend the equations of motion with the six trivial equations $\,dC_i/dt=\dot{C}_i\,$, we end up with nine first-order differential equations on the twelve variables
  $\,C_i(t)\,$, $\,\dot{C}_i\,$. Therefore, three arbitrary scalar constraints can be imposed on the said twelve variables.

  Deriving his variational equations, Lagrange set these three constraints so that the actual physical velocity in each point became equal to the Keplerian velocity (see Appendix A). Thereby he ensured that in each point of the orbit the instantaneous Keplerian conic (one defined by the instantaneous values of the time-varying orbital elements) is tangent to the orbit~--- so that the perturbed physical trajectory would coincide with the Keplerian orbit along which the body would move if the perturbing force were to cease instantaneously.  Written in the vector form, these constraints go under the name of \emph{osculation condition}. The corresponding instantaneous orbit is called an \emph{osculating orbit}, while the orbital elements satisfying the Lagrange constraint are called \emph{osculating orbital elements} \citep{kopeikin,seidelmann}.

  Mathematically, the imposition of any triple of scalar constraints (or of one vector constraint) confines the dynamics of the orbital state space to a 9-dimensional submanifold of the 12-dimensional manifold spanned by the orbital elements and their time derivatives \citep{efroim1,efroim2}. The choice of the constraints, however, is arbitrary. The employment of constraints different from Lagrange gives birth to \emph{non-osculating orbital elements}. Thus, while the physical orbit remains invariant to the particular selection of constraints, its description in the orbital-element space looks different for different constraint choices.

 Orbital elements, osculating or not, canonical or not, are useful mathematical variables that are not available in {\it{direct}} astronomical measurements.
 Direct measurements render us parallaxes, variations of brightness, etc. It is only {\it{a posteriori}} that we translate data into elements. At the same time,
 while not being immediately observable, elements may have convenient mathematical sense. For example, osculating elements parameterise a sequence of instantaneous conics sharing one focus and tangent to the physical orbit. Non-osculating elements parameterise instantaneous conics non-tangent to the physical orbit. However, non-osculating elements of a certain type (the so-called contact elements) osculate the orbit (i.e., are tangent to it) {\it{in the phase space}} \citep{efroim2,efroim3}. The choice of elements to employ is dictated by calculational convenience, analytical or numerical.

 With the Lagrange constraint relaxed, the Lagrange- or Delaunay-type orbital equations acquire the so-called {\it{gauge-generalised form}}.
 In this form, they are presented in \citep{efroim1,efroim2,efroim3,efroim4}. For an example of their employment in astrophysics, see the work by \citet{Dosopoulou}.
 In this paper, we derive a gauge-generalised form of the orbital equations in the form of Gauss.
 To this end, we utilise the freedom endowed by relaxing the Lagrange constraint, and arrive at new solutions to the relativistic PPN two-body problem of point masses. Fixing the gauge in three different ways, we find three such new solutions written in terms of non-osculating orbital elements. These solutions render orbits with either a fixed non-osculating semimajor axis, or a fixed non-osculating eccentricity, or a fixed non-osculating argument of periastron. We also develop a transformation from non-osculating to osculating orbital elements, and illustrate the new solutions using numerical simulations.

 \section{The Relativistic Force in the PPN Formalism}

 \subsection{The Lagrangian Perturbation and the Relativistic Force}

  When a Lagrangian $\;{\cal L}_0({\bf{\vec{r}}},\,{\bf{\dot{\vec{r
 }}}},\,t)\;=\;m\;{\bf{\dot{\vec{r}}}}^{\left.\,\right. 2}/2\,-\,U({\bf \vec
 r}\;,\;t)\,$ {is modified by an additional perturbing term} $\;\Delta{\cal{L}}({\bf{\vec{r}}},\,
 {\bf{\dot{\vec{r}}}},\,t)\;$, the Euler-Lagrange equations written for the
 perturbed setting
 \begin{eqnarray}
 {\cal L}({\bf{\vec r}},\,{\bf \dot{\vec r}},\,t)={\cal L}_0+
 \,\Delta {\cal L}\,=\,m\;\frac{~{\bf{\dot {\vec r}}}^{\left.\,\right. 2}}{2\;}\;-\;
 U({\bf \vec r},\,t)\,+
 \,\Delta {\cal L} ( {\bf \vec r},
 \,{ \bf { \dot { \vec {r}}}} ,\,t) \;\;
 \label{1}
 \end{eqnarray}
 acquire the form of:
 \begin{equation}
 m\;{\bf{\ddot { r}}}\;=\;-\;\frac{\partial U}{\partial {\bf{ r}}} \;+\;
 {\bf  F}\;\;\;\;,
 \label{2}
 \end{equation}
 the term ${\bf F}$ being the disturbing force:
 \ba
 {\bf  F}\;\equiv\;\frac{\partial \,\Delta {\cal L}}{\partial
 {\bf \vec r}}\;-\;\frac{d}{dt}\,\left(\frac{\partial \,\Delta {\cal L}}{\partial
 {\bf{\dot
 {\vec r}}}}\right)\;\;\;.\;
 \label{3}
 \ea
 We shall apply this formalism to the classical reduced two-body problem
 \ba
 {\cal{L}}_0({\bf{\vec{r}}},\,{\bf{\dot{\vec{r}}}})\;=\;m\;\frac{\,{\bf{\dot{\vec{r}}}}^{
 \left.\,\right. 2}}{2\;}\;+\;\frac{GM}{r}
 \label{4}
 \ea
 disturbed by the lowest-order relativistic correction {\citep{brumberg}}
 \ba
 \Delta{\cal L}\;=\;\frac{1}{c^2}\;\left[\;B_1\,\left(
 {\doterbold}^{\;{2}}\right)^2\;+
 \;B_2\;\frac{1}{r^2}\;+\;B_3\;\frac{{\doterbold}^{\;{2}}}{r}\;+\;
 B_4\;\frac{\left({\erbold\cdot\doterbold}\right)^{\;{2}}}{r^3}
 \right]\;\;\;,
 \label{5}
 \ea
 $B_1\;,\;.\,.\,.\;,\;B_4\;$ being constants {given by equation (\ref{6}) below}.

 Now $\,m\,$ has the meaning of the reduced mass:
 \ba
 m\,=\,\frac{m_1\;m_2}{m_1\,+\,m_2}\,\;,
 \label{}
 \ea
 while $\,M\,$ is the total mass of the system:
 \ba
 M\,=\,m_1\,+\,m_2\;\,,
 \label{}
 \ea
 $m_1$ and $m_2$ being the partners' masses.

 Insertion of (\ref{5}) into (\ref{3}) renders the so-called Chazy force. Historically it was pioneered by Chazy (1928) for General Relativity. We, however, shall be interested in a generalisation of this force to a broader class of theories, which fall within the scope of the PPN formalism.

 \subsection{The PPN Perturbation in the Chazy--Brumberg Form}

 Within the PPN parameterisation, which embraces both the GR and a class of its mathematical generalisations, the Lagrangian variation retains the above form, with the constants expressed through the Newton gravity constant $G$, the total mass $M$, and
 parameters $\,\alpha\,,\;\epsilon\,,\;\mu\,,\;\sigma\,$:
 \ba
 \nonumber
 &~&B_1\equiv\,\frac{2\alpha-2\epsilon+\mu}{8}~~~,\;~\;\;B_2\equiv (GM)^2
 \left(\epsilon+\,\frac{1}{2}\,\mu-\sigma\right)\,\;,\\
  \label{6}\\
 &~&B_3\equiv\,GM
 \left(\,-\,\alpha+\epsilon+\,\frac{1}{2}\,\mu\right)~~~,\;\;\;\;\;B_4\equiv
 GM\alpha\;\;\;,\;\;\;\;\;
 \nonumber
 \ea
 see equation (3.1.112) in \citep{brumberg}.

 An equivalent expression through $G$, $M$, and parameters $\,\alpha\,,\;\beta\,,\;\gamma\,$, as given by equation
 (3.1.45) from \citep{brumberg}, is in use also. Interrelation between the two parameterisations reads:
 \begin{subequations}
 \begin{eqnarray}
 \sigma~=~\beta~+~\gamma~-~\alpha~~~,~~~~~2\;\epsilon\;=\;\gamma\;+\;\alpha~~~,
 ~~~~~\mu\;=\;\gamma\;-\;\alpha\;+\;1~~~,
 \label{7a}
 \end{eqnarray}
 \mbox{the inverse formulae being}
 \begin{eqnarray}
 \alpha~=~\epsilon~-~\frac{1}{2}\;\mu~+~\frac{1}{2}~~~,~~~~~\beta\;=\;\sigma\;-\;
 \mu~+~1~~~,~~~~~\gamma\;=\;\epsilon\;+\;\frac{1}{2}\;\mu\;-\;\frac{1}{2}~~~.~~~~
 \label{7b}
 \end{eqnarray}
 \label{7}
 \end{subequations}

 Within the first parameterisation, the specific case of GR corresponds to
 \ba
 \sigma\;=\;2\;-\;\alpha~~~,~~~~~2~\epsilon~=~1~+~\alpha~~~,~~~~~
 \mu~=~2~-~\alpha\;\;\;.
 \label{9}
 \ea
 Within the second, to
 \ba
 \beta\;=\;\gamma\;=\;1\;\;\;.
 \label{8}
 \ea
 Consequently, the $B_i$ coefficients read, for GR, as
 \ba
 \nonumber
 &~&  B_1^{{^{(GR)}}}=\,\frac{1}{8}~~~,~~~~B_2^{{^{(GR)}}}=\,(GM)^2
 \left(\alpha\,-\,\frac{1}{2}\,\right)\;\,,\\
 \label{10}\\
 &~&  B_3^{{^{(GR)}}}=\,GM
 \left(\,-\,\alpha\,+\,\frac{3}{2}\,\right)~,~~~~~B_4^{{^{(GR)}}}\,=\,GM\,\alpha
 ~~.~~~~
 \nonumber
 \ea
 In GR, the PPN perturbation exerts on a {point-like} particle the so-called Chazy--Brumberg force, which assumes (see Appendix~B) the form
 \ba
 {\bf F}\;
 =\;\frac{GM}{c^2}\,\left[\;\left(\;\frac{2\,GM}{r}\;\,\sigma\,-\,
 2\;\epsilon\,(\,\doterbold\,)^{\,2}\,
 +\,3\,\alpha\,\frac{(\erbold\cdot\doterbold)^2}{r^2}\,\right)
 \;\frac{\erbold}{r^3}\,+\,2\,\mu\,
 \frac{(\erbold\cdot\doterbold)\,\doterbold}{r^3}\,\right]~~.~~~
 \label{11}
 \ea

  \subsection{The PPN Perturbation in the Damour-Deruelle Form}

 Another PPN-type parameterisation often used in the literature is the one suggested by \citet{dd}
  who employed parameters $\,\alpha_0\,$, $\,\alpha_1\,$, $\,\alpha_2\,$, $\,\beta_1\,$. The Chazy--Brumberg and Damour--Deruelle formulations are equivalent,
  as can be observed from the comparison of formulae (\ref{11}) and (\ref{drsw1}). The comparison yields the following link between the two parameterisations:
 \begin{subequations}
 \begin{eqnarray}
 \alpha_0(e)\;=\;2(\sigma\;-\;\epsilon)\;-\;(2\;\epsilon\;-\;2\;\mu\;-
 \;3\;\alpha)\;e^2\;=\;\;
 \;-\;2\;\alpha\;+\;2\;\beta\;+\;\gamma\;+\;(\gamma\;+
 \;2)\;e^2\;\;\;,\;\;\;\;\;\;\;
 \label{}\\
 \alpha_1\;~~\,=\;~~(2\;\sigma\;-\;4\;\epsilon )\;
   ~~~~~~~~~~~~~~~~~~~~~~~~~~~~~=\;\;
 \;2\;\beta\;-\;4\;\alpha\;\;\;,
 ~~~~~~~~~~~~~~~~~~~~~~~~~~~~~~~~~~~~
 \label{}\\
 \alpha_2\;~~\,=\;\,-\;(2\;\mu \;+\;3\;\alpha)\;
   ~~~~~~~~~~~~~~~~~~~~~~~~~~\,=\;\;
 \;-\;\alpha\;-\;2\;\gamma\;-\;2\;\;\;,\;\;\;\;\;\;\;\,
 ~~~~~~~~~~~~~~~~~~~~~~
 \label{}\\
 \beta_1 \,~~\,\;=\;~~2\;\mu\;
   \;\;\;\;\;\;\;\;\;\;\;~~~~~~~~~~~~~~~~~~~~~~~~~~~~~~~=\;\;
 \;2\;\gamma\;-\;2\;\alpha\;+\;2
 ~~~,~~~~~~~\,~~~~~~~~~~~~~~~~\;\;\;\;\;\;\;
 \end{eqnarray}
 \label{}
 \end{subequations}
 the inverse expressions being
  \begin{subequations}
 \begin{eqnarray}
 & &\alpha\;=\;-\;\frac{1}{3}\;(\alpha_2\;+\;\beta_1) \;\;\;,
 \label{}\\
 & &\mu\;=\;\frac{\beta_1}{2}\;\;\;,
 \label{}\\
 & &\epsilon\;=\;\frac{\alpha_0\;-\;\alpha_1\;+\;\alpha_2\;e^2}{2
     \;(1\,-\,e^2)}\;\;\;,
 \label{}\\
 & &\sigma\;=\;\frac{2\;\alpha_0\;-\;\alpha_1\;(1\;+\;e^2)\;+\;2\;
 \alpha_2\;e^2}{2\;(1\,-\,e^2)}\;\;\;,
 \label{}
 \end{eqnarray}
 \begin{eqnarray}
 & &~~~~~~~~~~~~~~~~~~\beta\,=\,\frac{2\,\alpha_0\,-\,\alpha_1\,(1\,+\,e^2)\,+\,2\,
 \alpha_2\,e^2}{2\,(1\,-\,e^2)}
 \,-\,\frac{1}{2}\,\beta_1\,+\,1\;\;,\;\qquad
 \label{}
 \end{eqnarray}
  \begin{eqnarray}
 & &\gamma\;=\;\frac{\alpha_0\;-\;\alpha_1\;+\;\alpha_2\;e^2}{2
     \;(1\,-\,e^2)}
 \;+\;\frac{1}{4}\;\beta_1\;-\;\frac{1}{2}\;\;\;.
 \label{}
 \end{eqnarray}
 \label{}
 \end{subequations}

\section{Gauge-invariant Gauss Variational Equations}

 Our goal in this section is to introduce a gauge-generalised form of the Gauss variational equations (GVEs). In this form, these equations will be used to derive new parameterisations of the relativistic two-body problem.

\subsection{Generalised Form of the Gauss Equations}

 We begin with a short reminder of how the gauge freedom (freedom of nonosculation) shows itself in celestial mechanics
 After this, we shall derive the gauge-generalised equations in the form of Gauss.

\subsubsection{Gauge freedom}

 To understand the underlying methodology, consider the equations of motion in the Newtonian two-body problem:
 \begin{equation}\label{eq1}
   \ddot{\vec r} + \frac{GM}{r^3}\vec r = \vec 0\;\;\,,
 \end{equation}
 where $M=(m_1+m_2)$ is the total mass, with $m_1$ and $m_2$ being the masses of the partners; and $\vec r$ is the position vector in some inertial frame of reference. The solution for $\vec r$ is known \citep{battin}, and can be symbolically expressed as
 \begin{equation}
 \label{eq2}
    \vec r = \vec f (t,\,C_1\,,\,\ldots\,,\,C_6)\;\;\;,
 \end{equation}
 where $C_i$ are constants of motion known as \emph{orbital elements}.

 Introducing the velocities ${\bf{\dot{r}}}$ and writing the equation of motion (\ref{eq1}) in the Cauchy form, we observe that (\ref{eq2})
 can be written as a mapping
  \be
  \; \left(\; C_{1} \, , \;...\; ,
\;C_{6}\;\right) \;\longleftrightarrow\; (\; x(t) \, , \; y(t)\, ,
\; z(t) \, , \; \dot{x}(t) \, , \; \dot{y}(t) \, , \; \dot{z}(t)
\; ) \;\;\;,
 \label{eq4a}
 \ee
  which is one-to-one over one orbit cycle, see Appendix A for details.

In the presence of a perturbing force $\vec F$, equation (\ref{eq1}) assumes the form
\begin{equation}\label{eq3}
   \ddot{\vec r} + \frac{GM}{r^3}\vec r = {\bf{F}}\;\,,
\end{equation}
 and its standard solution requires the ``constants'' of motion $C_i$ to become time-dependent:
 \begin{equation}\label{eq4}
    \vec r = \vec f (t,\,C_1(t),\ldots,C_6(t))\;\;\;.
 \end{equation}
 For time-dependent $\,C_i\,$, the velocity reads:
 \begin{equation}
 \label{eq5}
    \dot{\vec r} = \vec g\,+\,\vec\Phi\;\;\;,
 \end{equation}
 where
  \begin{equation}
 \label{eq9}
    \vec g(t,\,C_1,\ldots,C_6) \;\equiv\;  \frac{\p  }{\p t}\;{\vec{f}}(t,\,C_1,\ldots,C_6)\;\;\;,
 \end{equation}
 while the ``convective term'' is
 \begin{equation}
 \label{eq6}
    \vec\Phi \equiv \sum_{i=1}^6\frac{\p \vec f}{\p C_i}\dot C_i\;\;\;.
 \end{equation}
 The acceleration is then given by
 \footnote{~Be mindful that dot denotes a full time derivative. So, in equation (\ref{eq7}), the second term does {\it{not}} emerge as a partial time derivative of expression (\ref{eq6}). It shows up as a result of the differentiation of $\partial {{\bf {f}}}/\partial t$:
 $$
 \frac{d\,}{dt}\;\frac{\p \vec f}{\p t}\;=\; \frac{\p^2 \bf f}{\p t^2}+\sum_{i=1}^6\frac{\p ^2 \bf f}{\p C_i\p t }\dot C_i\;\;.
 $$
 }
 \begin{equation}
 \label{eq7}
    \ddot{\vec r} = \frac{\p^2 \bf f}{\p t^2}+\sum_{i=1}^6\frac{\p ^2 \bf f}{\p C_i\p t }\dot C_i+\dot {\vec \Phi}\;\;\;.
 \end{equation}
 As explained in Appendix A, correspondence (\ref{eq4a}) should now be changed to
 \ba
 \nonumber
  \;\left(\;C_1(t)\,,\; ...\; ,\; C_6(t)\,, \; \dot{C}_{1}(t)\,,\;
 ...\; ,\; \dot{C}_6(t)\;\right) \;\longrightarrow\\
  \label{eq4aa}\\
 (\;x(t)\,,\;y(t)\,,\;z(t)\,,\;\dot{x}(t)\,,\;\dot{y}(t)\,,\;\dot{z}(t)\;)
 \,\;,
\nonumber
 \ea
 a time-dependent mapping between a 12-dimensional and a 6-dimensional spaces. This mapping evidently cannot be one-to-one,
 and the resulting ambiguity of the parameterisation with $\,C_i(t)\,$ and $\,\dot{C}_i(t)\,$ can be removed by setting arbitrary constraints on the functions $\,C_i(t)\,$ and/or $\,\dot{C}_i(t)\,$. Since the insertion of ansatz (\ref{eq4}) into the three projections of the equation of motion (\ref{eq3}) furnishes us with three conditions on $\,C_i\,$ and $\,\dot{C}_i\;$ (equation (\ref{eq8}) below), only three arbitrary constraints are needed. This becomes especially clear when we write the equations of motion in the form of Cauchy.

 First observed in Efroimsky (\citeyear{efroim1}), this ambiguity is an internal symmetry, in that a gauge transformation (a switch from one set of constraints to another) leaves the physical trajectory unchanged. So, mathematically, it is analogous to the gauge freedom in electrodynamics.

 As was suggested in {\it{Ibid.}}, we do not set $\vec\Phi$ to be zero, but permit it to be an arbitrary vector function of the orbital parameters:
 \ba
 {\vec\Phi}\;=\;{\vec\Phi}(C_1(t)\,,\,...\,,\,C_6(t)\,)\;\;\;.
 \label{}
 \ea
 So we keep $\vec\Phi$ both in (\ref{eq5}) and in the subsequent derivations.
 Thereafter, we choose one or another particular functional form of $\,{\vec\Phi}(C_1(t)\,,\,...\,,\,C_6(t)\,)\,$ to nullify the rate of one or another orbital element, as given by the orbital equations.  Thus we employ the three projections of $\vec\Phi$ to set three arbitrary constraints whose necessity was discussed above. Stated alternatively, we use $\vec\Phi$ as a tool to remove the gauge freedom.
    An important caveat is that for ${{\vec{\Phi}}}\ne\vec 0$ the elements $C_i$ come out \emph{non-osculating}.

 Substituting expression (\ref{eq7}) for $\ddot{\vec r}$ into the equation of motion (\ref{eq3}), we obtain the following variational equations for the orbital elements:
 \begin{subequations}
 \label{eq8}
 \begin{eqnarray}
 && \sum_{i=1}^6\frac{\p \vec f}{\p C_i}\dot C_i = \vec \Phi\label{eq8a}\;\;\;,\\
 && \sum_{i=1}^6\frac{\p \vec g}{\p C_i}\dot C_i = \vec F-\dot{\vec \Phi}\label{eq8b}\;\;\,,
 \end{eqnarray}
 \end{subequations}
 where the first equation is a restatement of (\ref{eq6}),
 while $\vec g(t,\,C_1,\ldots,C_6)$ is the Keplerian velocity given by expression (\ref{eq9}).

  To combine equations (\ref{eq8}) into a single expression, we take the dot product of equation  (\ref{eq8a}) with $\p C_j/\p \vec f$ and equation  (\ref{eq8b}) with $\p C_j/\p \vec g$ and then add the two resulting expressions, arriving at
 \ba\label{eq10}
    \sum_{i=1}^6\left(\frac{\p C_j}{\p \vec f}\cdot\frac{\p \vec f}{\p C_i}+ \frac{\p C_j}{\p \vec g}\cdot\frac{\p \vec g}{\p C_i}\right)\dot {C}_i=\frac{\p C_j}{\p \vec g}\cdot(\vec F - \dot {\vec \Phi})+\frac{\p C_j}{\p \vec f}\cdot\vec\Phi
\ea
or, in vector form, \footnote{~An equivalent way to arrive at relation (\ref{eq11}) is to write
 \ba
 \nonumber
\dot{\vec C} = \frac{\p \vec C}{\p t}+\frac{\p\vec C}{\p \vec f}\left(\frac{\p \vec f}{\p t}+\frac{\p \vec f}{\p \vec C}\dot{\vec C}\right)+\frac{\p\vec C}{\p \vec g}\left(\frac{\p \vec g}{\p t}+\frac{\p \vec g}{\p \vec C}\dot{\vec C}\right)
 \ea
 and to recall that the analysis of the unperturbed problem rendered us
 \ba
 \nonumber
    \frac{\partial \vec C}{\partial t}\;=\;0\;\;\;,\quad
    \frac{\p \vec C}{\p \vec f}
    \;\frac{\p \vec f}{\p t}
  +
    \frac{\p \vec C}{\p \vec g}
    \;\frac{\p \vec g}{\p t}
    \;=\;0\;\,.
\ea
 Combined with one another and then with (\ref{eq8}), these formulae furnish us with (\ref{eq11}).
}
\begin{equation}\label{eq11}
 \dot {\vec C} \;=\; \frac{\p \vec C}{\p \vec g}\;(\vec F - \dot{\vec\Phi})\;+\;\frac{\p \vec C}{\p \vec f}\;\vec\Phi\;\,.
 \end{equation}
 Here we employed the notation $\vec C = [C_1,\ldots,C_6]^T$ and kept in mind that $C_i$ bears no explicit dependence on time.

 Pioneered in \citet{efroim1}, equations (\ref{eq11}) are the \emph{gauge-generalised} variational equations of orbital motion. The ensuing gauge-generalised orbital equations in the forms of Lagrange and Delaunay were derived by \citet{efroim3,efroim4}, see also \citet{efroim2}.  A detailed discussion on the topic is provided in the book by \citet{kopeikin}. For an example of practical use, see \citet{Dosopoulou}.

 \subsubsection{Derivation of the generalised Gauss-type equations}

 The derivation of the gauge-generalised orbital equations in the form of Gauss comprises two steps. First,
 it is convenient to keep on the right-hand side of (\ref{eq11}) only the partial time derivative of the gauge function $\vec\Phi$. To this end, we substitute
 \begin{equation}
 \label{manipulation}
    \vec{\dot\Phi}\;=\;\frac{\p\vec \Phi}{\p t}\;+\;\frac{\p\vec\Phi}{\p \vec C}\;\vec{\dot C}
 \end{equation}
 into the right-hand side of (\ref{eq11}), to obtain
 \begin{equation}\label{eqxx}
\left(I+\frac{\p\vec C}{\p \vec g}\frac{\p\vec \Phi}{\p \vec C}\right)\dot {\vec C} = \frac{\p \vec C}{\p \vec g}\left(\vec F - \frac{\p\vec \Phi}{\p t}\right)+\frac{\p \vec C}{\p \vec f}\vec\Phi\;\,,
\end{equation}
 where $I$ is a $6\times6$ identity matrix.

The second step is to express $\vec f$, $\vec g$, ${\p \vec C}/{\p \vec g}$, $\vec F$ and ${\p \vec C}/{\p \vec f}$ in a coordinate system with an origin at the primary. In this system, usually referred to as the $RSW$ frame, the unit vector $\hat{\vec R}$ is directed along the radius vector of the orbiter, radially outwards, while
$\hat{\vec S}$ is perpendicular to $\hat {\vec R}$,
residing in the instantaneous orbital plane defined by $\hat {\vec R}$ and the instantaneous orbital velocity. As ever,  $\hat{\vec W}$ completes the right-hand triad, so that
$\hat{\vec R}\times\hat{\vec S}=\hat{\vec W}$.

 Within the orthodox approach, $\vec\Phi$ and $\dot{\vec\Phi}$ appearing in equation  (\ref{eq11}) are set to be identically zero. This is why in all textbooks and papers, which develop the Gauss-type orbital equations, only the expression for ${\p \vec C}/{\p \vec g}$  are listed \citep{battin}. Thus, to obtain explicit expressions for the rates of change of the orbital elements, we shall have to first derive expressions for ${\p \vec C}/{\p \vec f}$, which are absent in the current literature. Presented in Appendix C, this derivation is performed by choosing the classical orbital elements as the phase variables, i.~e. $\vec C = [a,\,e,\,i,\Omega,\,\omega,\,l_0]^T$, where $a$ is the semimajor axis, $e$ is the eccentricity, $i$ is the inclination, $\Omega$ is the right ascension of the ascending node, $\omega$ is the argument of periastron, and $l_0$ is defined as \footnote{~To avoid confusion with our notation for the mass, we have used Brumberg's notation \citep{brumberg} for the mean anomaly.}
\begin{equation}\label{eq11a}
    l_0 = l-\int_{t_0}^t n dt\;\;\,,
\end{equation}
where $l$ is the mean anomaly, $t_0$ is a reference time (which may differ from the periastron passage time), and $n=\sqrt{GM/a^3}$ is the mean motion.

In addition, the position and velocity vectors in the $RSW$ frame are
\begin{equation}\label{eq12}
    [\vec r]_{
     \mathscr
    \mathbb{R}} = r\hat{\vec R},\,[\vec v]_{
     \mathscr
    \mathbb{R}} = \frac{\p r}{\p t}\hat{\vec R}+r \frac {\p f}{\p t}\hat{\vec S}\;\;\;,
\end{equation}
$f$ being the true anomaly. In the sequel, we omit the subindex $ R $ for brevity.

By collecting all the expressions derived in Appendix C and substituting them into equation  (\ref{eq11}), we obtain the gauge-invariant Gauss-type orbital equations. The resulting equations are valid in an arbitrary gauge, and are hence a generalised form of the classical Gauss-type equations. Writing $\vec F = F_R\hat{\vec R}+F_S\hat{\vec S}+F_W\hat{\vec W} = [F_R,\,F_S,\,F_W]^T$, $\vec\Phi = \Phi_R\hat{\vec R}+\Phi_S\hat{\vec S}+\Phi_W\hat{\vec W} = [\Phi_R,\,\Phi_S,\,\Phi_W]^T$ and $\dot{\vec\Phi} = \dot\Phi_R\hat{\vec R}+\dot\Phi_S\hat{\vec S}+\dot\Phi_W\hat{\vec W} = [\dot\Phi_R,\,\dot\Phi_S,\,\dot\Phi_W]^T$ provides the following gauge-generalised equations (wherein we used the notation $\dot \Phi \equiv \partial \Phi/\partial t$)

\begin{subequations}\label{gves}
\begin{eqnarray}
\frac{da}{dt} &=& 2\,{\frac { \left( { F_R}-{ \dot{\Phi}_R} \right) {a}^{2}e\,\sin
 f }{h}}+2\,{\frac { \left( { F_S}-{ \dot{\Phi}_S}
 \right) {a}^{2}p}{hr}}+2\,{\frac {{ \Phi_R}\,{a}^{2}}{{r}^{2}}} \;\;\;,
 \label{gvea}\\
\frac{de}{dt} &=&
{\frac { \left( { F_R}-{ \dot{\Phi}_R} \right) p\sin f }{h}}+{\frac { \left( { F_S}-{ \dot{\Phi}_S} \right)
 \left(  \left( p+r \right) \cos f +re \right) }{h}}\nonumber \\ &+&{
\frac {{ \Phi_R}\, \left( \cos f +e \right)  \left( 1+
e\,\cos f  \right) }{p}}+{\frac {{ \Phi_S}\,\sin
 f }{a}} \;\;\;,
 \label{gvee}\\
\frac{di}{dt} &=& {\frac { \left( { F_W}-{ \dot{\Phi}_W} \right) r\cos \left( f+\omega
 \right) }{h}}+{\frac {{ \Phi_W}\, \left( \sin \left( f+\omega \right) +e
\,\sin \omega  \right) }{p}} \;\;\;,
\label{gvei}\\
\frac{d\Omega}{dt} &=& {\frac { \left( { F_W}-{ \dot{\Phi}_W} \right) r\sin \left( f+\omega
 \right) }{h\sin i }}+{\frac {{ \Phi_W}\, \left[
 \left( \sin \left( f+\omega \right) +e\,\sin \omega  \right) r
\sin \left( f+\omega \right) -p \right] }{rp\sin i \cos
 \left( f+\omega \right) }} \;\;\;,
 \label{gveo}\nonumber\\ \\
\frac{d\omega}{dt} &=& -{\frac { \left( { F_R}-{ \dot{\Phi}_R} \right) p\cos f }{he}}+{\frac { \left( { F_S}-{ \dot{\Phi}_S} \right)
 \left( p+r \right) \sin f }{he}}\nonumber\\
 &-&{\frac { \left( {
F_W}-{ \dot{\Phi}_W} \right) r\sin \left( f+\omega \right) \cos   i
   }{h\sin i }}+{\frac {{ \Phi_R}\,\sin f  \left( 1+e\,\cos f  \right) }{pe}}-{\frac {{
 \Phi_S}\, \left( \cos f +e \right) }{pe}}\nonumber\\
 &+&{\frac {{
 \Phi_W}\, \left[  \left( \cos \left( f+\omega \right)  \right) ^{2}-\sin
 \left( f+\omega \right) e\,\sin \omega +e\,\cos f  \right] \cos i }{p\sin i \cos
 \left( f+\omega \right) }}  \;\;\;,
 \label{gvew}\\
\frac{d l_0}{dt}&=& \left( { F_R}-{ \dot{\Phi}_R} \right)  \left[  {\frac { \left( -2\,e+\cos f +e\,  \cos ^ 2 f  \right)  \left( 1-{e}^{2} \right) }{e\, \left( 1+e\,\cos f  \right) na}}  \right] \nonumber\\
&+& \left( { F_S}-{ \dot{\Phi}_S} \right)
 \left[ {\frac {   \left( e^2-1 \right)  \left( e\,
\cos f +2 \right) \sin f }{e\, \left( 1+
e\,\cos f  \right) na}} \right] \nonumber\\
&+&{ \Phi_R}\, \left[ {\frac {\ds-3-{e}^{2}+2\,   \cos^2 f(1+e^2)-2\,e\cos  f\sin^2f }{\ds 2\,a\,e\,\sqrt {1-{e}^{2}}\sin  f  }}  \right]\nonumber\\
&+&{\frac {
{ \Phi_S}\,\sqrt {1-{e}^{2}}\cos f }{a\,e}} \;\;\;,
\label{gvelam}
\end{eqnarray}
\end{subequations}
where $p=a(1-e^2)$ is the semi-latus rectum and $h=\sqrt{GM p}$ is
the magnitude of the angular momentum vector. Another useful
relationship is the gauge-generalised variational equation for the
true anomaly, obtained by using equation  (\ref{dfdr}):
\ba
\nonumber
    \dot f &=& \frac{h}{r^2}+\frac{1}{eh}\left[p\cos f  (F_R-\dot\Phi_R)-(p+r)\sin f  (F_S-\dot\Phi_S)\right]
    ~\\
    \label{fdot}\\
    &-& \frac{\Phi_R}{  ep\cos f}\sin ^ 2 f(1+e\cos f)+\frac{\Phi_S}{ep\cos f}\sin f(e^2\cos f+2e+\cos f)\;\;\;.
    \nonumber
\ea

%

 \subsection{The Post-Newtonian Force Model}
 The perturbing force in the first post-Newtonian approximation of general relativity can be written by using either the Chazy--Brumberg (\citeyear{chazy,brumberg}) or Damour -- Deruelle (\citeyear{dd}) forms. These two force models can be rendered equivalent by a proper choice of the force coefficients. We shall hence use the Damour -- Deruelle form.

Since the post-Newtonian force model bears direct dependance on the velocity, its parametrisations with osculating and non-osculating elements will be different. The expressions for the force components, written in the $RSW$ frame, are given by \citet{dd},
\begin{subequations}
\label{drsw0}
\begin{eqnarray}
F_R &=& \frac{GM}{c^2 r^2}\left[-(1+3\nu)\vec{\dot r}\cdot \vec{\dot r}+(4-0.5\nu)\frac{(\vec r\cdot\vec{\dot r})^2}{r^2}+(4+2\nu)\frac{GM}{r}\right]\\
F_S &=& \frac{GM}{c^2 r^3}(4-2\nu)(\vec r\cdot\vec{\dot r})  \vec{\dot r}\\
F_W &=& 0\;\,,
\end{eqnarray}
\end{subequations}
where $c$ is the speed of light, $M=m_1+m_2$ is the total mass, while $\nu$ denotes the dimensionless reduced mass:
 \ba
 \nu = \frac{m_1\,m_2}{(m_1+m_2)^2}\;\,\;.
 \label{}
 \ea

 Based on equation (\ref{eq5}), we have $\,\vec{\dot r} = \vec v+\vec \Phi\,$, with $\,\vec v$ given by equation (\ref{eq12}). Hence the gauge $\vec \Phi$ will show up in (\ref{drsw0}).

At this point, we shall perform an order-of-magnitude analysis in order to quantify the effect of the gauge on the force components, which is necessary for expressing the force components in terms of orbital elements. To that end, we define the dimensionless parameter
\begin{equation}\label{}
    \ve = \frac{GM}{c^2 p} \ll 1\;\;\,.
\end{equation}
We first note that the post-Newtonian force components can be expressed as a function of the small parameter $\ve$,
\begin{equation}
    \vec F = \ve \tilde{\vec F}\;\;\,.
\end{equation}
In addition, because the gauge $\vec \Phi$ can be determined \emph{ad lib}, we choose it to be proportional to $\ve$ as well, so that
\begin{equation}\label{gaugeeps}
     \vec \Phi = \ve\tilde{\vec \Phi}\;\;\,.
\end{equation}
Looking at \eqref{eqxx}, let us denote
\begin{equation}\label{eqlam}
    \Lambda \equiv \frac{\p\vec C}{\p \vec g}\frac{\p\vec \Phi}{\p \vec C}=\ve\tilde\Lambda\;\;\,,
\end{equation}
where, due to \eqref{gaugeeps},
\begin{equation}
     \tilde\Lambda \equiv \frac{\p\vec C}{\p \vec g}\frac{\p {\tilde{\vec \Phi}}}{\p \vec C}\;\;\,.
\end{equation}
We now utilise the approximation,
\begin{equation}
    \left(I+\ve \tilde{\Lambda}\right)^{-1}\approx I-\ve\tilde \Lambda,
\end{equation}
which is correct to first-order in $\ve$. We now re-write \eqref{eqxx} into
\begin{equation}
\dot {\vec C} = \left( I-\ve\tilde \Lambda\right)\left[\frac{\p \vec C}{\p \vec g}\left(\ve\tilde{\vec F} - \ve\frac{\p\tilde{\vec \Phi}}{\p t}\right)+\frac{\p \vec C}{\p \vec f}\ve\tilde{\vec\Phi}\right]\;\;\,.
\end{equation}
Retaining terms up to first-order in $\ve$ implies that equation (\ref{eqxx}) can be written as
\begin{equation}\label{cdotapprox}
\dot {\vec C} = \frac{\p \vec C}{\p \vec g}\left(\vec F - \frac{\p\vec \Phi}{\p t}\right)+\frac{\p \vec C}{\p \vec f}\vec\Phi\;\;\,.
\end{equation}

 Consequently, the gauge velocity in the presence of the post-Newtonian perturbations will be several orders of magnitude smaller than the nominal Keplerian orbital velocity.
 (in our example considered later on, $\mathscr O(\ve)=10^{-8}$; for e.g. binary pulsars, it can be larger, though still much smaller than unity.)
 Therefore, since we are seeking first-order solutions, we can omit second-order expressions of the gauge; this will result in the following expressions for the force components, written in terms of non-osculating orbital elements:
\begin{subequations}
\label{drsw1}
\begin{eqnarray}
    F_R &=& \frac{GM \ve}{ p^2}(1+e\cos f)^2\left[\alpha_0(e)+\alpha_1 e\cos f+\alpha_2 e^2\cos^2 f\right]\nonumber\\
        &-&\sqrt{\frac{GM}{p^3}}\frac{\ve\Phi_R}{2}(1+e\cos f )^2(-12 +14\nu) e \sin f \nonumber\\
        &-&\sqrt{\frac{GM}{p^3}}\frac{\ve\Phi_S}{2}(1+e\cos f )^2[(12\nu  +4)e\cos f +4+12\nu]\;\;\,,\\[1.5ex]
    F_S &=& \frac{GM \ve}{ p^2}(1+e\cos f)^3\beta_1 e\sin f\nonumber \\
        &+& \sqrt{\frac{GM}{p^3}}\ve\Phi_R\beta_1(1+e\cos f)^3\nonumber\\
        &+& \sqrt{\frac{GM}{p^3}}\ve\Phi_S\beta_1(1+e\cos f)^2e\sin f\;\;\,,\\[1.5ex]
    F_W &=& 0\;\;\,,
\end{eqnarray}
\end{subequations}
where
\begin{eqnarray}\label{coef}
    &&\alpha_0(e) = 3-\nu+3e^2-3.5\nu  e^2,\,\alpha_1 = 2-4\nu,\,\alpha_2 = -4+0.5\nu,\,\nonumber\\&&\beta_1 = 4-2\nu = 0.4(\alpha_1-2\alpha_2)\;\;\,.
\end{eqnarray}
To first order, neglecting the terms $\ve\Phi_R$ and $\ve\Phi_S$, we obtain the expressions
\begin{subequations}
\label{drsw}
\begin{eqnarray}
    F_R &=& \frac{G^2M^2}{c^2p^3}(1+e\cos f)^2\left[\alpha_0(e)+\alpha_1 e\cos f+\alpha_2 e^2\cos^2 f\right]\;\;\,,\\
    F_S &=& \frac{G^2M^2}{c^2p^3}(1+e\cos f)^3\beta_1 e\sin f\;\;\,,\\
    F_W &=& 0\;\;\,,
\end{eqnarray}
\end{subequations}
which are identical to the force components written in terms of the osculating elements. Thus, to first order, the gauge velocity does not affect the post-Newtonian perturbation.

In the relativistic reduced two-body problem, we have $\,\nu\to 0\,$; and the coefficients become
\begin{equation}\label{coef1}
    \alpha_0(e) = 3+3e^2\;\,,\;\;\;\alpha_1 = 2\;,\;\;\;\alpha_2 = -4\;,\;\;\;\beta_1 = 4\;\,.
\end{equation}
\section{Fixing the Gauge}
Both Brumberg (\citeyear{brumberg}) and Damour and Deruelle (\citeyear{dd}) solved the GVEs (\ref{gves}) with osculating elements (i.e.  $\,\vec\Phi = \vec 0\,$ and $\,\dot{\vec \Phi} = \vec 0$) and the post-Keplerian force components (\ref{drsw}). The solution was obtained under the following assumptions:

\begin{enumerate}
\item \label{ass1}The orbital elements appearing in the right-hand side of the GVEs are treated as constants.
\item \label{ass2}The independent variable is transformed from time to true anomaly by approximating (cf. \citep{brumberg}, p.~7;  {in this approximation, we use only the Keplerian rate of the true anomaly})
\begin{equation}\label{}
    \frac{df}{dt}\approx\frac{\p f}{ \p t} = \frac{h}{r^2} = \sqrt{\frac{GM}{p^3}}(1+e\cos f)^2
\end{equation}
so that, using the notation $(\cdot)'$ to denote derivative with respect to true anomaly,
\begin{equation}\label{}
    \dot{C}_j \approx C_j' \frac{\p f}{\p t}\;\;\,.
\end{equation}

\end{enumerate}
Assumptions \ref{ass1} and \ref{ass2} imply that the solutions obtained for the osculating orbital elements under the post-Keplerian perturbation are \emph{first-order approximations}; nevertheless, these solutions are of prime importance for both modeling and experimental validation of general relativity.

To obtain \emph{new} {parameterisations}, we shall solve the GVEs (\ref{gves}) with \emph{non-osculating} elements, i.~e. $\vec\Phi$ and $\dot{\vec\Phi}$ will \emph{not} be identically zero. In fact, we shall use the extra degrees-of-freedom introduced by $\vec\Phi$ to nullify the rates of change of some of the orbital elements, thus providing simpler parameterisations of the relativistic two-body problem. To that end, we shall adopt Assumptions \ref{ass1} and \ref{ass2}, and, in the spirit of Assumption \ref{ass2}, write that (cf. equation (\ref{cdotapprox}))
\begin{equation}\label{}
   \frac{\p\vec\Phi}{\p t}  = \frac{\p\vec\Phi}{\p f}\frac{\p f}{\p t} = \frac{h}{r^2} \vec\Phi'\;\;\,.
\end{equation}
This will allow us to use $f$ as an independent variable in a seamless manner.

In addition, since the post-Newtonian effect {manifests itself as a perturbing force} in the orbital plane ($F_W = 0$), $\Omega$ and $i$ remain constant. To preserve the two-dimensionality of the orbit parametrisation with non-osculating elements, we must therefore set $\Phi_W=\dot\Phi_W=0$. We are thus left with two degrees-of freedom -- $(\Phi_R,\,\dot\Phi_R;\Phi_S,\dot{\Phi}_S)$ -- that can be used to nullify, at most, two additional orbital elements. We shall now derive sample closed-form solutions illustrating the utility of the proposed approach.

\subsection{Constant Semimajor Axis}

 Looking at equations (\ref{gvea}) and (\ref{gvee}), we note that there could be potentials $\Phi_R$ and $\Phi_S$ found that simultaneously nullify $\dot a$ and $\dot e$. However, this would require solving two linear non-autonomous, non-homogenous differential equations. We shall thus nullify $\dot a $ and $\dot e$ separately.

From equation  (\ref{gvea}) it is evident that if $\Phi_R \equiv 0$, then $\dot a=0$ if
 \begin{equation}\label{phisdot1}
    \Phi_S' =\frac{G^2 M^2 e \sin f [\alpha_0+\alpha_1e\cos f +\alpha_2e^2\cos ^ 2 f+\beta_1(1+e\cos f)^2]}{c^2 \sqrt{GM p^3}(1+e\cos f)}
 \end{equation}
and, for some integration constant $\kappa_1$,
\begin{eqnarray}\label{phis1}
    \Phi_S &=& \int\Phi_S'\,df \nonumber\\ &=& -\frac{G^2M^2    [2(\alpha_0-\alpha_1+\alpha_2)\ln(1+e\cc_ f)+[(\alpha_2+\beta_1)e^2\cc_f^2+2(\alpha_1-\alpha_2+\beta_1)]e\cc_ f ]}{2c^2 \sqrt{GM p^3}}\nonumber\\&+&\kappa_1\;\;\,,
\end{eqnarray}
where we have used the compact notation $\cc_x=\cos x,\,\s_x=\sin x$. The resulting semimajor axis now satisfies $a = a_0$, where $(\cdot)_0$ denotes an initial value, and the remaining orbital elements can be evaluated by substituting equations (\ref{phisdot1}) and (\ref{phis1}) into the GVEs (\ref{gves}).

{Brumberg (\citeyear{brumberg}, p.~88) introduced a modified semimajor axis $a^\star$, expressed as a function of the semimajor axis and eccentricity. In our presentation, the non-osculating semimajor axis equals the initial semimajor axis.}


A useful approximation for $\Phi_S$ given in equation  (\ref{phis1}) can be obtained for near-circular motion in the restricted two-body problem; in this case,
\begin{equation}\label{}
    \Phi_S =  -\frac{7e}{c^2}\left(\frac{GM}{a}\right)^{\frac{3}{2}}\cos f +
 \mathscr
O(e^2)
\end{equation}
and the approximate post-Newtonian parametrisation with non-osculating elements is given by
\begin{subequations}
\begin{eqnarray}
a &=& a_0\;\;\,,\\
e &=& e_0\left[1+\frac{GM}{4c^2a}\left(11 -12\cc_ f/e_0 +11  \cc_{ 2f}\right)\right]+
  \mathscr
O(e^2)\;\;\,,\\
\omega &=& \omega_0+\frac{GM}{4c^2 a}\left[-2f-(15e+12e^{-1})\s_ f+11\s_{ 2f}-7e\s_{ 3f}\right]+
  \mathscr
 O(e^2)\;\;\,,\\
l_0 &=& (l_0)_0+\frac{GM}{4c^2a}\left[ (12e^{-1}+45e)\s_f -11(\s_{2f}+2f)+7e\s_{3f}\right]+
  \mathscr
 O(e^2)\;\;.\;\;\qquad
\end{eqnarray}
\end{subequations}

\subsection{Constant Eccentricity}
Requiring that $\dot e=0$ and choosing $\Phi_S\equiv 0$ yields the differential equation
\begin{eqnarray}\label{}
&&-\frac{\s_ f(1+e\cc_ f)^2}{p}\Phi_R(f)'+\frac{(\cc_ f+e)(1+e\cc_ f)}{p}\Phi_R(f)=\frac{G^2 M^2(1+e\cc_f )^2\s_ f}{hp^2c^2}\nonumber\\
&\cdot& \left\{\alpha_0+\alpha_1e\cc_f+\alpha_2 e^2\cc_f^2+\beta_1e [(2+e\cc_f)\cc_f+e]\right\}\;\;\,,
\end{eqnarray}
whose solution is given by
\begin{eqnarray}
\label{phire}
 \Phi_R(f) &=& \frac{G^2M^2\s_ f}{2c^2\sqrt{GM p^3}(1+e\cc_ f)}\left\{\ln\left[\frac{2\s_ f}{\cc_ f(1+\cc_ f)}\right][(4\beta_1 +2\alpha_0+2\alpha_1) e+(2\alpha_2+4\beta_1 )e^3] \right.\nonumber\\
 &+&\left.\left[(\beta_1+\alpha_2 )e(1-\cc_f^2) -(6\beta_1 +2\alpha_2+2\alpha_1)(\cc_ f+1)\right]\right.
 \nonumber\\
 &-&\left. \ln\left[\frac{ 1-\cc_ f}{\s_ f} \right]\left[ (2\alpha_2+8\beta_1+2\alpha_1)e^2+2\alpha_0\right]\right\}\;\;\,.
\end{eqnarray}
The resulting eccentricity satisfies $e = e_0$; the remaining elements can be calculated using the GVEs (\ref{gves}).

Similarly to the previous procedure, a simple approximation for $\Phi_R$ (\ref{phire}), linear in the eccentricity, can be obtained for near-circular motion in the restricted relativistic two-body problem:
\begin{equation}\label{}
    \Phi_R \approx \frac{\s _f}{c^2}\left(\frac{GM}{a}\right)^{\frac{3}{2}}\left\{ 3\ln\left(\frac{1-\cc_f}{\s_f}\right)(e\cc_ f-1)+10(e\cc_f^2-1)+13e\ln\left[\frac{2\s_f}{\cc_f(1+\cc_f)}\right]\right\}\;\;\,.
\end{equation}

\subsection{Constant Argument of Periastron}

The secular growth of $\omega$, which is linearly proportional to $f$ using the osculating elements (this is consistent in all the representations listed by \citet{kk}), can be nullified using non-osculating elements and a particular gauge choice. Requiring that $\dot\omega=0$ and choosing $\Phi_S\equiv 0$ yields the following differential equation:
\begin{eqnarray}\label{odephi}
  &&\frac{\cos f(1+e\cos f)^2}{pe}\Phi_R'(f)+\frac{\sin f(1+e\cos f)}{ p e} \Phi_R(f)
=\frac{G^2M^2(1+e\cos f)^2}{h p^2 c^2}  \nonumber\\ &\cdot&\left[-\frac{(\alpha_0+\alpha_1 e\cos f+\alpha_2 e^2 \cos^2f )\cos f}{e}+(2+e\cos f)b_1 \sin ^2 f\right]
\end{eqnarray}
Equation (\ref{odephi}) can be solved for $\Phi_R(f)$, yielding
\begin{eqnarray}\label{phirw}
\Phi_R(f) &=& \frac{1}{2}\frac{G^2M^2}{hc^2p(1+e\cc_f)}\left\{(-6\beta_1 e^2\cc_f+2\alpha_0\cc_f)\ln(\sec f+\tan f)\right.
\nonumber\\&+&\left.[(\alpha_2-\beta_1) e^2   +(2 \alpha_1+4 \beta_1+2 \alpha_0)]e\cc_f\,f\right.\nonumber\\
&+&\left.\s_ f[(\beta_1+\alpha_2)e^3\cc^2_f +(2 \alpha_1+2\alpha_2+6\beta_1)e^2\cc_f   -4\beta_1 e ]\right\}\nonumber\\&+&\frac{\kappa_2\cc_f}{(1+e\cc_f)}\;\;\,.
\end{eqnarray}
where $\kappa_2$ is a constant of integration. Choosing $\Phi_R$ as in equation  (\ref{phirw}) will yield $\omega=\omega_0$. The remaining elements can be calculated using the GVEs (\ref{gves}).

We can again obtain a simple approximation for $\Phi_R$ by assuming near-circular motion in the restricted two-body setup:
\begin{equation}\label{}
    \Phi_R \approx \frac{1}{c^2}\left(\frac{GM}{a}\right)^{\frac{3}{2}}\left[ 3\cc_ f\ln\left(\frac{1+\s_ f}{\cc_ f}\right)(1-e\cc_ f)+e(13f\cc_ f-8\s_ f)\right]\;\;\,.
\end{equation}
\subsection{The Relation Between the Osculating and Non-osculating Elements}

An important aspect of the ongoing discussion is finding a relationship between the osculating elements and the non-osculating elements. It is important to keep in mind that the non-osculating elements do not possess the same physical meaning of their osculating counterparts. Thus, for analytical studies as well as for numerical exploitation of the new parameterisations offered above, expressions relating the osculating and non-osculating elements must be provided. These relations will be later used for a numerical illustration of the new formalism.

To begin, let us denote an osculating element by $C_i^\star$. For the purpose of this discussion, it would be convenient to use the true anomaly as one of the orbital element; thus, a set of osculating elements that uniquely determine the position and velocity in inertial space are $\vec C^\star =\{a^\star,\,e^\star,\,i^\star,\,\Omega^\star,\,\omega^\star,\,f^\star\}$, and the corresponding set of non-osculating elements is $\vec C =\{a,\,e,\,i,\,\Omega,\,\omega,\,f\}$. Because the position remains invariant to a selection of a gauge,
\begin{equation}\label{rr}
    \vec r(\vec C^\star) = \vec r(\vec C)\;\;\,.
\end{equation}
The relationship between the velocities is given by
\begin{equation}\label{vv}
    \vec g(\vec C^\star) = \vec g(\vec C )+\vec\Phi(\vec C )\;\,.
\end{equation}
Equations (\ref{rr}) and (\ref{vv}) define six algebraic equations, that can be solved for $ \vec C^\star$.

For the post-Newtonian perturbation, the motion is two dimensional, since the node and inclination remain unaffected by the presence of the relativistic correction. Thus, without loss of generality, we can choose $i=i^\star=\Omega=\Omega^\star \equiv 0$. In this case, the position vector assumes the form
\begin{equation}\label{rosc}
    \vec r(\vec C) = \frac{a(1-e^2)}{1+e\cos f}\left[\cos(f+\omega),\,\sin(f+\omega),\,0 \right]^T
\end{equation}
and the Keplerian velocity is
\begin{equation}\label{vosc}
    \vec g(\vec C) = \sqrt{\frac{GM  }{a(1-e^2) }}\left[ -e\sin \omega -\sin (f+\omega),\,e\cos \omega +\cos (f+\omega),\,0 \right]^T\;\;\,.
\end{equation}
The algebraic equations ensuing from (\ref{rr}) and (\ref{vv}) are quite complex and require numerics.

\section{Numerical Example}
Consider, for illustration purposes, the relativistic correction of Mercury's orbit about the Sun. In osculating elements, using the subscript 0 to denote initial values, the orbit is given by the initial conditions (in a J2000 system of reference)
\begin{equation}\label{}
    a_0^\star = 57910000\, \km,\,e_0^\star = 0.2056,\,\omega_0^\star = 1.351870079406362\,\rad,\,f_0^\star = 0\;\;\,.
\end{equation}
The problem parameters are
\begin{equation}\label{icosc}
    GM = 1.327120220308192\cdot 10^{11}\, \km^3/\sec^2,\,\nu = 1.660046706402425\cdot 10^{-7}\;\;\,.
\end{equation}
The non-osculating initial elements, obtained by solving equations (\ref{rr}) and (\ref{vv}), are
\begin{eqnarray}\label{icnosc}
    &&a_0 = 57910001.278597308147\, \km,\,e_0  = 0.20560001753959287403\nonumber\\
    &&\omega_0  = 1.351869887693436\,\rad,\,f_0  = 0.19171292543848736670\cdot 10^{-6}\,\rad\;\;\,.\nonumber\\
\end{eqnarray}

Equations (\ref{gves}), (\ref{fdot}) can be integrated with the initial conditions (\ref{icosc}), to yield the time history of the osculating elements, and initial conditions (\ref{icnosc}), with a given gauge function, to yield the time history of the non-osculating elements.

\begin{center}
\begin{figure}[htbp!]
    \includegraphics[width = .9\linewidth]{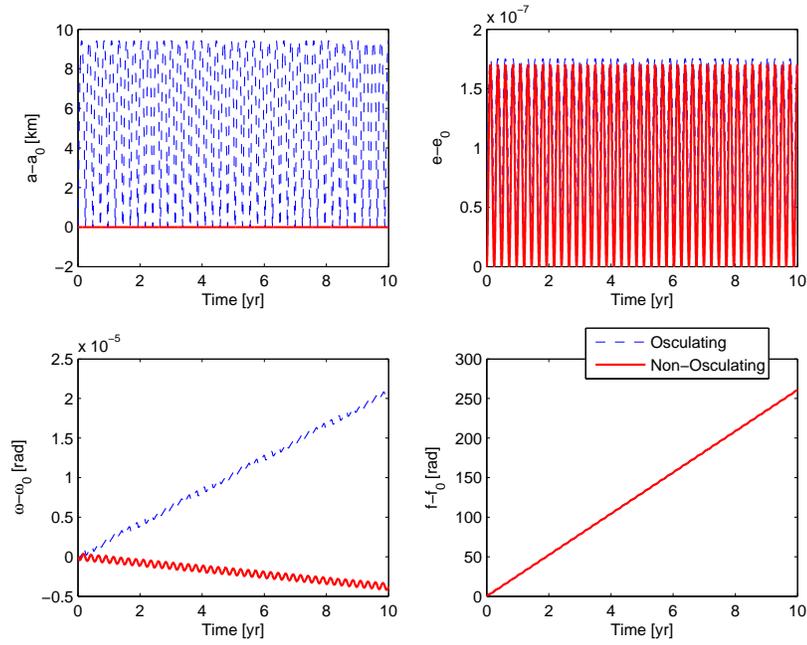}
    ~\\ \vspace{25mm}
    \caption{A comparison between the osculating and non-osculating elements. The non-osculating semimajor axis remains fixed due to the effect of the gauge velocity. The drift of the argument of periastron is much smaller.}
    \label{fig:elements}
\end{figure}
\end{center}

We shall provide a comparison of the time histories of the osculating and non-osculating elements for the gauge (\ref{phis1}), yielding a constant semimajor axis, with $\kappa1$ arbitrarily chosen as zero. This comparison is depicted in Figure \ref{fig:elements}, showing the osculating and non-osculating semimajor axis, eccentricity, argument of periastron and true anomaly. The initial values of the semimajor axis and eccentricity were subtracted from the time histories of these elements to accentuate the post-Newtonian effect. As expected, the non-osculating semimajor axis remains constant, while the osculating semimajor axis oscillates, reaching a maximum of about 9.5 km. The eccentricity in both cases behaves similarly. The argument of periastron drifts linearly (on average) in both cases, although the rotation of the non-osculating apsidal line is retrograde. The differences between the anomalies are two small to show on the given scale.

\begin{center}
\begin{figure}[htbp!]
    \includegraphics[width = .9\linewidth]{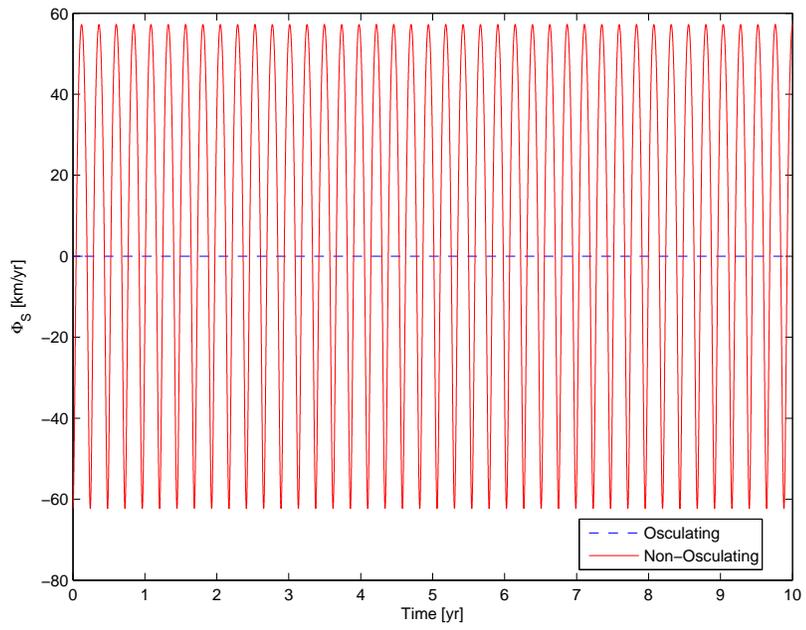}
    ~\\ \vspace{25mm}
    \caption{The time history of the gauge velocity. This velocity is about 7 orders of magnitude smaller than the nominal orbital velocity.}
    \label{fig:gauge}
\end{figure}
\end{center}

Figure \ref{fig:gauge} shows the time history of the gauge function, which is zero in the osculating case. In the non-osculating case, the gauge velocity oscillates between the values of $\pm60$ km/yr. This value is about 7 orders of magnitude smaller than the nominal orbital velocity.

\section{Conclusions}

Utilising the generalised Lagrange constraint for deriving the Gauss variational equations introduces gauge freedom, which can be utilised to derive new solutions for orbital motion affected by a post-Netwonian perturbation. The gauge velocity required to derive the new solutions is a few orders of magnitude smaller than the nominal orbital velocity. Whereas the classical PPN solutions entail variation of all six classical osculating orbital elements, the transformation to non-osculating elements portrays a different picture, wherein some of the elements can be kept constant. This is achieved by relaxing the traditional requirement to represent a perturbed two-body orbit by a series of Keplerian conics instantaneously osculating to the perturbed trajectory.
As opposed to known solutions of the post-Newtonian perturbation, wherein the mean eccentricity and mean semimajor axis remain constant, and the argument of periastron exhibits secular growth -- the solutions developed in this paper provide exact nullification of the time-varying eccentricity, or the semimajor axis, or the argument of periastron, without requiring averaging, and without utilising a canonical transformation to switch between osculating and mean elements.

\section*{Appendix A.\\
 Transition from Cartesian to orbital coordinates}

 Referring the reader for detailed treatment to the preprint by \citet{efroim1} and review \citep{efroim2}, we provide here a squeezed explanation of why the switch from Cartesian to orbital coordinates produces internal freedom.

 The reduced two-body problem,
 \ba
 \label{1001}
{{\bf {\ddot { {r}}}}}\;+\;\frac{G(m_{1}\,+
 \,m_{2})}{r^2}\; \frac{{{\bf
r}}}{r}\;=\;0\;\;\;,
\qquad \erbold\;\equiv\;\erbold_{2}\,-\;\erbold_{1}
 \;\;\;,
  \ea
 has a generic solution, a Keplerian conic, which can be expressed, in some fixed Cartesian frame, as
  \ba
 \nonumber x\;=\;f_1\left(C_1, ... , C_6, \,t
 \right)\;\;\;,\;\;\;\;\;\;\;\;\;\;\dot x\;=\;{\rm g}_1\left(C_1, ... , C_6,
 \,t \right)\;\;\;\;,\;\\ y\;=\;f_2\left(C_1, ... , C_6, \,t
 \right)\;\;\;,\;\;\;\;\;\;\;\;\;\;\dot y\;= \;{\rm g}_2\left(C_1, ... ,
 C_6, \,t \right)\;\;\;\;,\;
 \label{1002}\\
 \nonumber z\;=\;f_3\left(C_1, ... , C_6, \,t
 \right)\;\;\;,\;\;\;\;\;\;\;\;\;\;\dot z\;= \;{\rm g}_3\left(C_1, ... ,
 C_6, \,t \right)\;\;\;\;,\;\,
  \ea
   or, shortly:
    \be
    \erbold \;=\; {\bf
  f} \left(C_1, ... , C_6, \,t \right)\;\;\;,\;\;\;\;\;\;
 \;\;\;\;{\bf \dot{\erbold}} \;=\;{\bf  g}\left(C_1, ... , C_6,
 \,t \right) \;\;\;\;\;\;.\;\;\;\;\;\;
 \label{1003}
 \ee
 Here $C_i$ are six adjustable constants, while ${\rm g}_i$ are the partial derivatives of $f_i$ over the last argument:
 \be
 {{\bf{{g}}}}\left(C_1, ... , C_6, \,t \right)\;\equiv\;\left(\frac{\partial {\bf { f}}}{
 \partial
 t}\right)_{C=const} \;\;\;.
 \label{1004}
 \ee

 Now consider a perturbed problem, where the disturbing force $\,\Delta{\bf{F}}\,$ is an arbitrary vector-valued function of the position and velocity:
 \be
 {\bf \ddot{
r}}\;+\;\frac{\mu}{r^2}\;\frac{{\bf r }}{r}\;=\;{\bf{ F}}\;\;\;.
 \label{1008}
 \ee
 Following
 \citet{Lagrange_1808a,Lagrange_1808b,Lagrange_1809},
 we utilise solution (\ref{1002} - \ref{1004}) as an ansatz for a solution
 to the perturbed problem (\ref{1008}), the ``constants" now being time dependent:
 \be
{\bf  r }\;=\;{\bf  f} \left(C_1(t), ... , C_6(t), \,t
\right)\,\;\;\;,
 \label{1009}
 \ee
 and the functional form of $\,\bf{\vec{f}}\,$ being the same as in (\ref{1003}). The velocity
 \be
 \frac{d \erbold}{dt}\;=\;\frac{ \partial {\bf
f}}{\partial t}\;+\; \sum_i \;\frac{\partial {\bf
f}}{\partial C_i}\;\frac{d C_i}{d t}\;= \;{\bf g}\;+\; \sum_i
\;\frac{\partial {\bf  f}}{\partial C_i}\;\frac{d C_i}{d
t}
 \label{10010}
 \ee
 now contains a ``convective'' term $\;\sum ({\partial{\bf f}}/{\partial C_i})({dC_i}/{dt})\;$, while the first term,
 $\;\bf{{g}}\;$, has the same functional form as in (\ref{1004}).

 The insertion of $\; {\bf f} \left(C_1(t), ... , C_6(t), \,t \right)\,$ into the perturbed equation of motion (\ref{1008}) renders three scalar differential equations of the second order. They contain an independent parameter, time, and six time-dependent quantities $\,C_i( t)\,$
 whose evolution is to be found. This cannot be carried out in a unique way because the number of variables exceeds, by three, the number of equations. So, while the actual physical orbit (comprising the locus of points in the Cartesian coordinates and the values of velocity in each point) is unique, its parametrisation through the orbital variables $\,C_i(t)\,$ is ambiguous.
 Lagrange chose to remove this freedom by setting three independent constraints on $\,\dot{C}_i(t)\;$:
  \be
  \sum_i \;\frac{\partial
{\bf{\vec{f}}}}{\partial C_i}\;\frac{d C_i}{d t}\;=\;0 \;\;\;,
   \label{3.15}
  \ee
 These constraints nullify the ``convective'' term in expression (\ref{10010}) and thereby make the velocity equal to the Keplerian velocity $\,{\bf g}\,$. Stated alternatively, these constraints make the instantaneous conic tangent to the physical orbit.

 Within the unperturbed two-body problem (\ref{1001}), its solution (\ref{1003}) defines a time-dependent one-to-one (within one
 orbital period) mapping
 \be
  \; \left(\; C_{1} \, , \;...\; ,
\;C_{6}\;\right) \;\longleftrightarrow\; (\; x(t) \, , \; y(t)\, ,
\; z(t) \, , \; \dot{x}(t) \, , \; \dot{y}(t) \, , \; \dot{z}(t)
\; ) \;\;\;.
 \label{3.5}
 \ee
 Under disturbance, ansatz (\ref{1009}) entails the emergence of the time derivatives $\,\dot{C}_i\,$ in (\ref{10010}). So, instead of (\ref{3.5}), we obtain a time-dependent mapping between a 12-dimensional and a 6-dimensional spaces:
 \ba
 \nonumber
  \;\left(\;C_1(t)\,,\; ...\; ,\; C_6(t)\,, \; \dot{C}_{1}(t)\,,\;
 ...\; ,\; \dot{C}_6(t)\;\right) \;\longrightarrow\\
  \label{3.8}\\
 (\;x(t)\,,\;y(t)\,,\;z(t)\,,\;\dot{x}(t)\,,\;\dot{y}(t)\,,\;\dot{z}(t)\;)
 \,\;.
\nonumber
 \ea
 This mapping cannot be one-to-one. Its ambiguity is another expression of the obvious fact that the three equations of motion (\ref{1008}) are insufficient to determine the six functions $\,C_1\,,\,...\,,\,C_6\,$, for which reason one can impose three arbitrary constraints on these functions and their derivatives.\,\footnote{ The dynamics, in the form of first-order differential
  equations for the orbital coordinates $\;C_i(t)\;$ and their derivatives $\;H_i(t )\,\equiv\,{\bf{\dot{\rm{C}}}}_i(t)\;$,
  will include six evident first-order identities for these twelve functions: $\;H_i(t)\;=\;dC_i(t)/dt\;$. Three more differential
  equations will be obtained by substituting $\,\erbold = \efbold (C_1,...,C_6,t)\,$ into (\ref{1008}). These equations will be of the
  second order in $\;C_i(t)\;$. However, in terms of both $\;C_i(t)\;$ and $\;H_i(t)\;$ these equations will be of the first order only.
  Altogether, we have nine first-order equations for twelve functions $\;C_i(t)\;$ and $\;H_i(t)\;$. Hence, the problem is underdefined
  and permits three extra conditions to be imposed by hand. \label{footnote}
  }
 Together with the three equations of motion, the three constraints remove the freedom and make mapping (\ref{3.8}) unambiguous.

 From the mathematical point of view, the Lagrange constraint (\ref{3.15}) is completely arbitrary. We could as well have chosen some different
 supplementary condition
 \be
\sum_i \;\frac{\partial {\bf \vec f}}{\partial C_i}\;\frac{d
C_i}{d t}\;=\; {\bf {\vec \Phi}}(C_{1\,,\,...\,,\,6}\,,\,t)\;\;\;,\;
 \label{10013}
  \ee
 with $\Phibold\,$ an arbitrary function of time and the parameters $\;C_i\;$.\,\footnote{~$\Phibold\,$ can be imparted also with a dependence on the parameters' time derivatives of all orders. Then higher-than-first-order time derivatives of $\,C_i\,$ will show up in the subsequent developments. This will imply that additional initial conditions, beyond those on ${\bf\vec r}$ and ${\bf\dot{\vec r}}$, will have to be specified in order to close the system. To avoid this complication, we set $\Phibold\,$ to be a function of time and $\,C_i\,$ only.}
 The arbitrariness of these conditions reveals the ambiguity of the
  representation of an orbit by instantaneous Keplerian ellipses. Mappings between different representations reveal an internal symmetry
  (and a symmetry group) underlying this formalism.

 Substitution of (\ref{3.15}) with (\ref{10013}) leaves the physical motion unchanged, but alters its mathematical description. Specifically, it
 entails different solutions for the orbital elements. This invariance of a physical theory under a change of parametrisation goes under the names of gauge symmetry. The gauge transformations make a group, which is isomorphic to the group of all real-valued functions on $\mathbb R^3$, with the group operation being addition \citep{gurfilaip}.

 Just like in electrodynamics, a right choice of gauge often simplifies the solution of the equations of motion. A deliberate choice of non-osculating orbital elements~--- i.e., of a set $\,C_i\,$ obeying some condition (\ref{10013}) different from (\ref{3.15})~--- can simplify the expressions for these elements' rates.

\section*{Appendix B. Derivation of the Generalised Chazy--Brumberg Force\\
}

 Through the medium of the formulae
 \ba
 \frac{\partial ~}{\partial \erbold}\;\frac{1}{r^n}\;=\;-\;n\;\frac{\erbold}{r^{n+2}}\;
 \;\;~~~~~~
 \mbox{and}~~~~~~~~
 \frac{\partial ~}{\partial \erbold}\;\left({\erbold\cdot\doterbold}\right)^{\;{2}}\;=\;
 2\;\left({\erbold\cdot\doterbold}\right)\;\doterbold\;\;\;,
 \label{A1}
 \ea
 we derive from (\ref{5}) the following:
 \ba
 \frac{\partial \Delta \cal L}{\partial\erbold}\;=\;\frac{1}{c^2}\left[\,-\,2\;\frac{B_2
 }{r^4}\;\erbold\,-\,B_3\,\frac{\doterbold^{\;2}}{r^3}\;\erbold\;-\;3\;B_4\;\frac{(
 \erbold\cdot\doterbold)^{\;2}}{r^5}\;\erbold\;+\;2\,B_4\,(\erbold\cdot\doterbold)\,
 \frac{\doterbold}{r^3}\;\right]\;\;\;.
 \label{A2}
 \ea
 Also, using
 \ba
 \frac{\partial\doterbold^{\;2}}{\partial \doterbold}\;=\;2\;\doterbold\;\;\;~,
 ~~~~~
 \frac{\partial\left(\doterbold^{\;2}\right)^{\,2}}{\partial \doterbold}\;=\;4\;
 (\,\doterbold\,)^{\,2}\;\doterbold\;\;\;~,~~~~
 \mbox{and}~~~~~~
 \frac{\partial(\erbold\cdot\doterbold)^{\,2}}{\partial\doterbold}\;=\;2\;
 \left(\erbold\cdot\doterbold\right)\;\erbold~~~~,~~~
 \label{A3}
 \ea
 we deduce from (\ref{5}) that
 \ba
 \frac{\partial \Delta\cal L}{\partial \doterbold}\;=\;\frac{1}{c^2}\,\left[4\,B_1\,(\,
 \doterbold\,)^{\,2}\,\doterbold\,+\;\frac{2\,B_3}{r}\,\doterbold\,+\;\frac{2\,B_4
 }{r^3}\,(\erbold\cdot\doterbold)\,\erbold\,\right]\;\;\;.
 \label{A4}
 \ea
 The subsequent calculations, though elementary, need care. For $\;\doterbold\;
 \partial^2 \Delta{\cal L}/\partial \erbold\,\partial\doterbold\;$, we have:
 \ba
 \nonumber
 \doterbold\;\frac{\partial^2 \Delta{\cal L}}{\partial \erbold\;\partial\doterbold}\;=
 \;\frac{1}{c^2}\;\sum_i\,\dot{r}^i\;\frac{\partial}{\partial r^i}\,\left[\,4\,B_1\,(\,
 \doterbold\,)^{\,2}\,\doterbold\,+\;\frac{2\,B_3}{r}\,\doterbold\,+\;\frac{2\,B_4
 }{r^3}\,(\erbold\cdot\doterbold)\,\erbold \,\right]\;=
 \ea
 \ba
 \nonumber
 \frac{1}{c^2}\;\sum_i\,\left[\,2\,B_3\,\left(\;-\;\frac{\dot{r}^i\,r^i}{r^3}\,
 \right)\,\doterbold\;+\;2\,B_4\,\left(\;-\;3\;\frac{\dot{r}^i\,r^i}{r^5}\,
 \left(\,\erbold\,\cdot\,\doterbold\,\right)\,\erbold\,\right)\;+\;
 \frac{2\,B_4}{r^3}\,\left(\;{\dot{r}^i}\;{\dot{r}^i}\,\right)\,\erbold\;+\;
 \frac{2\,B_4}{r^3}\;{\dot{r}^i}\left(\,\doterbold\,\cdot\,\erbold\,\right)\;
 \hat{\bf{e}}_i\;\,\right]
 \ea
 \ba
 =\;\frac{1}{c^2}\;\left[\;-\;2\;B_3\;\frac{\erbold\,\cdot\,\doterbold}{r^3}\;\,\doterbold
 \;-\;6\;B_4\;\frac{\left(\,\erbold\,\cdot\,\doterbold\,\right)^2}{r^5}\;\,\erbold\;
 +\;2\;B_4\;\frac{\doterbold^{\;2}}{r^3}\;\,\erbold\;+\;\frac{2\;B_4}{r^3}\;\left(\,\erbold\,\cdot\,
 \doterbold\,\right)\;\doterbold\,\right]\;\;\;.\;\;\;\;\;\;\;\;\;\;\;
 \label{A5}
 \ea
 By similar means, for $\;\ddoterbold\,
 \partial^2 \Delta{\cal L}/\partial \doterbold\partial\doterbold\;$ we obtain:
 \ba
 \nonumber
 \ddoterbold\;\frac{\partial^2 \Delta{\cal L}}{\partial \doterbold\;\partial\doterbold}\;=
 \;\frac{1}{c^2}\;\sum_i\,\ddot{r}^i\;\frac{\partial}{\partial \dot{r}^i}\,\left[\,4\,B_1\,(\,
 \doterbold\,)^{\,2}\,\doterbold\,+\;\frac{2\,B_3}{r}\,\doterbold\,+\;\frac{2\,B_4
 }{r^3}\,(\erbold\cdot\doterbold)\,\erbold \,\right]\;=
 \ea
 \ba
 \nonumber
 \frac{1}{c^2}\;\sum_i\,\left[\;\ddot{r}^i\;4\;B_1\;2\;\dot{r}^{\,i}\;\doterbold\;+\;\ddot{r}^i\;4\;B_1\;\left(\,\dot{r}\,
 \right)^2\;\hat{\bf{e}}_i\;+\;\ddot{r}^i\;\frac{2\;B_3}{r}\;\hat{\bf{e}}_i\;+\;
 \ddot{r}^i\;\frac{2\;B_1}{r^3}\;\dot{r}^i\;\erbold\;\right]
 \ea
 \ba
 =\;\frac{1}{c^2}\;\left[\;8\;B_1\;\left(\,\ddoterbold\,\cdot\,\doterbold\,\right)\;
 \doterbold\;+\;4\;B_1\;\ddoterbold\;\left(\,\doterbold\,\right)^2\;+\;
 \ddoterbold\;\frac{2\;B_3}{r}\;+\;\left(\,\ddoterbold\,\cdot\,\erbold\,\right)\;\frac{2\;B_4}{r^3}\;\erbold
 \;\right]\;\;\;.\;\;\;\;
 \label{A6}
 \ea
 Together, (\ref{A5}) and (\ref{A6}) yield:
 \ba
 \nonumber
 -\;\doterbold\;\frac{\partial^2 \Delta{\cal L}}{\partial \erbold\;\partial\doterbold}\;
 -\;\ddoterbold\;\frac{\partial^2 \Delta{\cal L}}{\partial \doterbold\;\partial\doterbold}
 \,=\,
 -\,\frac{1}{c^2}\,\left[\;-\,2\,B_3\,\frac{\erbold\,\cdot\,\doterbold}{r^3}\;\,\doterbold
 \;-\;6\;B_4\;\frac{\left(\,\erbold\,\cdot\,\doterbold\,\right)^2}{r^5}\;\,\erbold\;
 +\;2\;B_4\;\frac{\doterbold^{\;2}}{r^3}\;\,\erbold\;+\;\frac{2\;B_4}{r^3}\;\left(\,\erbold\,\cdot\,
 \doterbold\,\right)\;\doterbold\,\right]
 \ea
 \ba
 \nonumber
 -\;\frac{1}{c^2}\;\left[\;8\;B_1\;\left(\,\ddoterbold\,\cdot\,\doterbold\,\right)\;
 \doterbold\;+\;4\;B_1\;\ddoterbold\;\left(\,\doterbold\,\right)^2\;+\;
 \frac{2\;B_3}{r}\;\ddoterbold\;+\;\left(\,\ddoterbold\,\cdot\,\erbold\,\right)\;\frac{2\;
 B_4}{r^3}\;\erbold
 \;\right]
 \ea
 \ba
 \nonumber
 =\;-\;\frac{\doterbold}{c^2}\;\left[\;8\;B_1\;\left(\,\ddoterbold\,\cdot\,\doterbold\,\right)\;
 -\;2\;B_3\;\frac{\erbold\,\cdot\,\doterbold}{r^3}\;+\;\frac{2\;B_4}{r^3}\;\left(\,\erbold\,\cdot\,
 \doterbold\,\right)\;\right]~~~~~~~~~~~~~~~~~~~~~~~~~~~~~~~~~~~~~~~~
 ~~~~~~~~~~~~~~~\\
 \label{A7}\\
 \nonumber
 -\;\frac{\erbold}{c^2}\;\left[\;
 -\;6\;B_4\;\frac{\left(\,\erbold\,\cdot\,\doterbold\,\right)^2}{r^5}\;
 +\;2\;B_4\;\frac{\doterbold^{\;2}}{r^3}\;+\;\left(\,\ddoterbold\,\cdot\,\erbold\,\right)\;\frac{2\;
 B_4}{r^3}
 \;\right]\;
 -\;\frac{\ddoterbold}{c^2}\;\left[\;4\;B_1\;\left(\,\doterbold\,\right)^2\;+\;
 \frac{2\;B_3}{r}\;\right]~~~.~~~
 \ea
 Making use of
 \ba
 \ddoterbold\;=\;-\;\frac{GM}{r^3}\,\erbold\,+\,\frac{GM}{r^3}\,\frac{1}{c^2}\,\left[
 \;.\;.\;.\;\right]\;\approx\;-\;\frac{GM}{r^3}\,\erbold\;\;\;,
 \label{A8}
 \ea
 we then arrive at
 \ba
 \nonumber
 -\;\doterbold\;\frac{\partial^2 \Delta{\cal L}}{\partial \erbold\;\partial\doterbold}\;
 -\;\ddoterbold\;\frac{\partial^2 \Delta{\cal L}}{\partial \doterbold\;\partial\doterbold}
 \;=\;-\;\frac{\doterbold}{c^2}\;\left[\;-\;8\;B_1\;\frac{GM}{r^3}\;\left(\,\erbold\,\cdot\,\doterbold\,
 \right)\;-\;2\;B_3\;\frac{\erbold\,\cdot\,\doterbold}{r^3}\;+\;\frac{2\;B_4}{r^3}\;\left(\,\erbold\,
 \cdot\,\doterbold\,\right)\;\right]
 \ea
 ~\\
 \ba
 -\;\frac{\erbold}{c^2}\;\left[\;
 -\;6\;B_4\;\frac{\left(\,\erbold\,\cdot\,\doterbold\,\right)^2}{r^5}\;
 +\;2\;B_4\;\frac{\doterbold^{\;2}}{r^3}\;-\;\frac{GM}{r^3}\;\left(\,\erbold\,\cdot\,\erbold\,\right)\;\frac{2\;
 B_4}{r^3}
 -\;\frac{GM}{r^3}\;\left(\;4\;B_1\;\doterbold^{\;2}\;+\;
 \frac{2\;B_3}{r}\;\right)\;\right]~~~.~~~
 \label{A9}
 \ea
 Combined, formulae (\ref{3}), (\ref{A2}), and (\ref{A9}) entail:
 \ba
 \nonumber
 {\bf{F}}\;=\;\frac{\partial \Delta {\cal L}}{\partial \erbold}\;-\;{\doterbold}\,\;\frac{
 \partial^2 \Delta {\cal L}}{\partial\erbold\,\partial {\stackrel{~}{\doterbold}}}\;-\;
 \ddoterbold\,\;\frac{\partial^2 \Delta\cal L}{\partial^2 {\stackrel{~}{
 \doterbold}}}\;\approx~~~~~~~~~~~~~~~~~~~~~~~~~~~~~~~~~\\
 \nonumber\\
 \nonumber\\
 \nonumber\\
 \nonumber
 \frac{1}{c^2}\left[\,-\,2\;\frac{B_2}{r^4}\;\erbold\,-\,B_3\,\frac{\doterbold^{\;2}}{
 r^3}\;\erbold\;-\;3\;B_4\;\frac{(\erbold\cdot\doterbold)^{\;2}}{r^5}\;\erbold
 +\;2\;B_4\;\left(\,\erbold\,\cdot\,\doterbold\,\right)\;\frac{\doterbold}{r^3}
 \;\right]
 \ea
 ~\\
 \ba
 \nonumber
 -\;\frac{\doterbold}{c^2}\;\left[\;-\;8\;B_1\;\frac{GM}{r^3}\;\left(\,\erbold\,\cdot\,\doterbold\,
 \right)\;-\;2\;B_3\;\frac{\erbold\,\cdot\,\doterbold}{r^3}\;+\;2\;B_4\;
 \frac{\erbold\,
 \cdot\,\doterbold}{r^3}\;\right]
 \ea
 ~\\
 \ba
 \nonumber
 -\;\frac{\erbold}{c^2}\;\left[\;
 -\;6\;B_4\;\frac{\left(\,\erbold\,\cdot\,\doterbold\,\right)^2}{r^5}\;
 +\;2\;B_4\;\frac{\doterbold^{\;2}}{r^3}\;-\;\frac{GM}{r^3}\;\left(\,\erbold\,\cdot\,\erbold\,\right)\;\frac{2\;
 B_4}{r^3}
 -\;\frac{GM}{r^3}\;\left(\;4\;B_1\;\doterbold^{\;2}\;+\;
 \frac{2\;B_3}{r}\;\right)\;\right]~~~~~~
 \ea
 ~\\
 \ba
 \nonumber
 =\;\frac{1}{c^2}\,\left\{\,\frac{2}{r^4}\;\erbold\;\left[\,GM\,(B_3\,+\,B_4)\,-\,B_2
 \right]\,+\,
 \left[\;-\,B_3\,+\,4\,GM\,B_1\,-\,2\,B_4\,\right]\,\frac{(\,\doterbold\,)^{\,2}}{r^3}\;\erbold\,+
 \,\frac{3\,B_4}{r^5}\,(\erbold\cdot\doterbold)^2\,\erbold
 \right. \\
 \nonumber\\
 \nonumber\\
 \left.
 +\;\frac{(\erbold\cdot\doterbold)\,\doterbold}{r^3}\;\left[\;8\;B_1\;G\;M\;
 +\;2\;B_3\;\right]\;\,\right\}~~~.~~~~~~~~~~~~~~
 \label{A10}
 \ea
 A further substitution of (\ref{6}) results in
 \ba
 \nonumber
 {\bf F}\,=\,
 \frac{1}{c^2}\,\left[\frac{2\sigma}{r^4}\,(GM)^2\,\erbold\,+\,GM\,
 (\,-\,2\;\epsilon\,)\,\frac{(\,\doterbold\,)^{\,2}}{r^3}\,\erbold\,
 +\,\frac{3\,GM\,\alpha}{r^5}\,(\erbold\cdot\doterbold)^2\,\erbold\,+\,2\,\mu\,
 \frac{(\erbold\cdot\doterbold)\,\doterbold}{r^3}\,\right]
 \ea
 \ba
 =\;\frac{GM}{c^2}\,\left[\;\left(\;\frac{2\,GM}{r}\;\,\sigma\,-\,
 2\;\epsilon\,(\,\doterbold\,)^{\,2}\,
 +\,3\,\alpha\,\frac{(\erbold\cdot\doterbold)^2}{r^2}\,\right)
 \;\frac{\erbold}{r^3}\,+\,2\,\mu\,
 \frac{(\erbold\cdot\doterbold)\,\doterbold}{r^3}\,\right]~~.~~~
 \label{A11}
 \ea


%
%

 \section*{Appendix C. Calculation of $\p\vec C/\p\vec r$}

 In this Appendix, all orbital elements are non-osculating. The known Keplerian relations are evaluated so that $v$ is dependent upon the non-osculating elements. In accordance with equation (\ref{eq4}), we set
 \ba
 {\vec r}\;\equiv\;{\vec f}(t,\,C_1,\,...\,,\,C_6)\;\;\;.
 \label{}
 \ea
 while by $\vec v$ we understand the Keplerian velocity along an instantaneous (non-osculating) Keplerian ellipse,
 i.e. the same as $\,{\vec g}(t,\,C_1\,,\,...\,,\,C_6)\,$ in equations (\ref{eq5} - \ref{eq9}):
  \ba
  {\vec v}\;\equiv\;{\vec g}(t,\,C_1,\,...\,,\,C_6)\;\;\;.
  \label{}
  \ea
  Under this premise, we can use all known expressions for Keplerian conics. E.g., to calculate $\p a/\p \vec r$, we write the vis-viva equation
 \begin{equation}
 \label{eq13}
    \frac{v^2}{2}-\frac{GM}{r} = -\frac{GM}{2a}\;\;\;.
 \end{equation}
 Differentiation of this expression with respect to $\vec r$ yields
 \begin{equation}
 \label{eq14}
    \frac{\p a}{\p \vec r} = 2a^2\frac{\vec r}{r^3}\;\;\;,
 \end{equation}
 which in the polar representation reads as
 \begin{equation}
 \label{eq15}
    \frac{\p a}{\p \vec r} = \left[\frac{2a^2}{r^2},\,0,\,0\right]\;\;\;.
 \end{equation}
 The next step is to evaluate the partial derivative of $h$, the magnitude of the Keplerian angular momentum vector $\vec h$. To that end, we use the definition
 \begin{equation}\label{eq16}
    \vec h = \vec r \times \vec v = -S_v\vec r\;\;\;,
 \end{equation}
 where $S_v$ is the matrix cross-product equivalent,
 \begin{equation}
 \label{eq17}
    S_v = \left[
            \begin{array}{ccc}
              0 & 0 & r\frac{\textstyle\p f}{\textstyle\p t} \\
              0 & 0 & -\frac{\textstyle\p r}{\textstyle\p t} \\
              -r\frac{\textstyle\p f}{\textstyle\p t} & \frac{\textstyle\p r}{\textstyle\p t} & 0 \\
            \end{array}
          \right]\quad.
 \end{equation}
 Using $h^2=\vec h^T\vec h$, we deduce that
 \begin{equation}
 \label{eq18}
    2h\frac{\p h}{\p \vec r} = 2\vec h^T\frac{\p\vec h}{\p \vec r} =-2\vec h^T S_v\;\;\;,
 \end{equation}
 wherefrom
 \begin{equation}
 \label{eq19}
    \frac{\p h}{\p \vec r}= -\hat{\vec h} S_v =-\hat{\vec W} S_v = \left[r\frac{\p f}{\p t},\,-\frac{\p r}{\p t},\,0\right]\;\;\;.
 \end{equation}
 We now proceed to obtain $\p e/\p\vec r$. This is performed by differentiating both sides of the relation $h=\sqrt{GM a (1-e^2)}\;$:
 \begin{equation}
 \label{eq20}
    \frac{\p h}{\p \vec r}=-\frac{GM}{2h}\left[(1-e^2)\frac{\p a}{\p\vec r}-2ae\frac{\p e}{\p \vec r}\right]\;\;\;.
 \end{equation}
 At this point, it is useful to introduce the Keplerian formulae
 \begin{equation}\label{eq21}
    r = \frac{p}{1+e\cos f}
 \end{equation}
 and
 \begin{equation}\label{eq22}
    \frac{\p f}{\p t} = \frac{h}{r^2}=\sqrt{\frac{GM}{p^3}}(1+e \cos f)^2
 \end{equation}
 where $p=a(1-e^2)=h^2/(GM)$ is the semilatus rectum. Equation (\ref{eq22}) can be used to transform the partial time derivatives to partial derivatives with respect to the true anomaly, a procedure essential for facilitating the underlying algebra. We have:
 \begin{equation}\label{eq23}
    \frac{\p r}{\p t} = \frac{\p r}{\p f} \frac{\p f}{\p t}  = \frac{e h \sin f}{r(1+e\cos f)}\;\;\;.
 \end{equation}
 The employment of equations (\ref{eq15}), (\ref{eq19}), and (\ref{eq23}) furnishes us with
 \begin{equation}\label{eq24}
    \frac{\p e}{\p \vec r} = \left[\frac{(\cos f+e)(1+e\cos f)}{p},\,\frac{\sin f}{a},\,0\right]\;\;\;.
 \end{equation}
 A similar procedure may be exploited to find $\p i/\p \vec r$. We start out with the definition
 \begin{equation}\label{eq25}
    \cos i = \frac{\vec{h}^T\hat{\vec k}}{h}\;\;\;,
 \end{equation}
 where $\hat{\vec k}$ is a unit vector normal to the fundamental plane of an inertial reference system, whose components in the $RSW$ frame are given by \citep{battin}
 \begin{equation}\label{eq26}
 \hat{\vec k} = [\sin i\sin(f+\omega),\,\sin i \cos(f+\omega),\,\cos i]^T\;\;\;.
 \end{equation}
 Differentiation of equation  (\ref{eq25}) with respect to $\vec r$ entails
 \begin{equation}\label{eq27}
    \sin i\frac{\p i}{\p \vec r} = \frac{1}{h}\left(\frac{\p h}{\p \vec r}\cos i -S_v\hat{\vec k}\right)\;\;\;.
 \end{equation}
 By using the expressions for $S_v$, $\p h/\p \vec r$ and $\hat{\vec k}$ given by equations (\ref{eq17}), (\ref{eq19}) and (\ref{eq26}), respectively, while keeping in mind equations (\ref{eq22}) and (\ref{eq23}), one arrives at
 \begin{equation}\label{eq28}
    \frac{\p i}{\p \vec r} = \left[0,\,0,\,\frac{e\sin(f+\omega)\sin\omega}{p}\right]\;\;\;.
 \end{equation}
 At this point, one may proceed in one of the following ways: using the definitions of the remaining orbital elements, take the partial derivatives with respect to $\vec r$ as in the preceding procedure; or, using the known results for the derivatives of the elements with respect to $\vec v$ and their Poisson brackets, $(C_i,\,C_j)$, solve for the remaining unknown derivatives of the elements with respect to $\vec r$. We shall adopt the second option. Thus, to obtain e.g. $\p \Omega/\p\vec r$, one can solve the equations
 \begin{subequations}
 \begin{eqnarray}
 \label{eq29}
    \frac{\p i}{\p \vec r}\left(\frac{\p\Omega}{\p \vec v}\right)^T-\frac{\p i}{\p \vec v}\left(\frac{\p\Omega}{\p \vec r}\right)^T&=&(i,\,\Omega)\;\;\,.\\
    \frac{\p a}{\p \vec r}\left(\frac{\p\Omega}{\p \vec v}\right)^T-\frac{\p a}{\p \vec v}\left(\frac{\p\Omega}{\p \vec r}\right)^T&=&(a,\,\Omega)\;\;\,.\\
       \frac{\p e}{\p \vec r}\left(\frac{\p\Omega}{\p \vec v}\right)^T-\frac{\p e}{\p \vec v}\left(\frac{\p\Omega}{\p \vec r}\right)^T&=&(e,\,\Omega)\;\;\,.
 \end{eqnarray}
 \end{subequations}
 where \citep{battin}
 \begin{equation}\label{eq30}
    (a,\,\Omega)=(e,\,\Omega)=0,\,(i,\,\Omega)=\frac{1}{h\sin i}
 \end{equation}
and
\begin{eqnarray}\label{eq31}
 && \frac{\p a}{\p \vec v} = \frac{2a^2}{h}\left[e\sin f,\,\frac{p}{r},\,0\right],\,\frac{\p e}{\p \vec v}=\frac{1}{h}\left[p\sin f,\,(p+r)\cos(f)+re,\,0\right] \nonumber  \\
 &&\frac{\p i}{\p \vec v}=\frac{1}{h}[0,\,0,\,r\cos(f+\omega)]\;\;\,.
\end{eqnarray}
Solving equations (\ref{eq29}) for $\p \Omega/\p\vec r$, we obtain:
\begin{equation}\label{eq32}
    \frac{\p \Omega}{\p\vec r} = \left[0,\,0,\,\frac{[\sin(f+\omega)+e\sin\omega] r\sin(f+\omega)-p}{rp\sin i\cos(f+\omega)}\right]\;\;\,.
\end{equation}
Similarly, the expression for $\p\omega/\p\vec r$ can be derived by solving the equations
\begin{subequations}
\begin{eqnarray}\label{eq33}
    \frac{\p \Omega}{\p \vec r}\left(\frac{\p\omega}{\p \vec v}\right)^T-\frac{\p \Omega}{\p \vec v}\left(\frac{\p\omega}{\p \vec r}\right)^T&=&(\Omega,\,\omega)\;\;\,,\\
    \frac{\p a}{\p \vec r}\left(\frac{\p\omega}{\p \vec v}\right)^T-\frac{\p a}{\p \vec v}\left(\frac{\p\omega}{\p \vec r}\right)^T&=&(a,\,\omega)\;\;\,,\\
       \frac{\p e}{\p \vec r}\left(\frac{\p\omega}{\p \vec v}\right)^T-\frac{\p e}{\p \vec v}\left(\frac{\p\omega}{\p \vec r}\right)^T&=&(e,\,\omega)\;\;\,,
\end{eqnarray}
\end{subequations}
where \citep{battin}
\begin{equation}\label{eq34}
    (a,\,\omega)=(\Omega,\,\omega)=0,\,(e,\,\omega)=\frac{h}{aeGM}
\end{equation}
and
\begin{eqnarray}\label{eq35}
 && \frac{\p \Omega}{\p \vec v} = \left[0,\,0,\,\frac{r\sin(f+\omega)}{h\sin i}\right]  \nonumber  \\
 &&\frac{\p \omega}{\p \vec v}= \frac{1}{h}\left[-\frac{p\cos f}{e},\,\frac{(p+r)\sin f}{ e},\,-\frac{r\sin(f+\omega)\cos i}{\sin i}\right]\;\;\,.
\end{eqnarray}
This procedure provides the expression
\begin{equation}\label{eq36}
    \left(\frac{\p\omega}{\p\vec r}\right)^T =\left[
                                 \begin{array}{c}
                                   \frac{\ds \sin f(e\cos f+1)}{\ds pe}  \\[1.3ex]

                                    -\frac{\ds e+\cos f}{\ds pe}\\[1.3ex]
 \frac{\ds [\cos^2(f+\omega)-\sin(f+\omega)e\sin(\omega)+e\cos f]\cot i }{\ds p\cos(f+\omega) }\\[1.3ex]
                                 \end{array}
                               \right]\;\;\,.
\end{equation}
The expression for $\p l_0/\p\vec r$ can be derived by solving the equations
\begin{subequations}
\begin{eqnarray}\label{eq37}
    \frac{\p \Omega}{\p \vec r}\left(\frac{\p l_0}{\p \vec v}\right)^T-\frac{\p \Omega}{\p \vec v}\left(\frac{\p l_0}{\p \vec r}\right)^T&=&(\Omega,\, l_0)\;\;\,,\\
       \frac{\p a}{\p \vec r}\left(\frac{\p l_0}{\p \vec v}\right)^T-\frac{\p a}{\p \vec v}\left(\frac{\p l_0}{\p \vec r}\right)^T&=&(a,\, l_0)\;\;\,,\\
       \frac{\p e}{\p \vec r}\left(\frac{\p l_0}{\p \vec v}\right)^T-\frac{\p e}{\p \vec v}\left(\frac{\p l_0}{\p \vec r}\right)^T&=&(e,\, l_0)\;\;\,,
\end{eqnarray}
\end{subequations}
where \citep{battin}
\begin{equation}\label{eq38}
    (\Omega,\, l_0)=0,\,(a,\, l_0)=\frac{2}{na},\,(e,\, l_0)=\frac{-1+e^2}{e\sqrt{aGM}}
\end{equation}
and
\begin{equation}\label{eq39}
  \left(\frac{\p  l_0}{\p \vec v}\right)^T =  \left[
                                                   \begin{array}{c}
                                                      {\frac {\ds \left( -2\,e+\cos  f   +e   \cos ^2 f
    \right)  \left(  1-{e}^{2} \right) }{\ds e \left( 1
+e\cos   f    \right) na}}
 \\[3.0ex]
                                                     {\frac {\ds \left( e^2-1 \right)    \left( 2+e\cos
   f   \right) \sin f   }{\ds e \left( 1+e\cos
   f    \right) na}}
 \\ \\[1.5ex]
                                                     0
                                                   \end{array}
                                                 \right]\;\;\,.
\end{equation}
This results in
\begin{equation}\label{eq40}
   \left( \frac{\p l_0}{\p\vec r} \right)^T = \left[
                                                                        \begin{array}{c}
                                                                         {\frac {\ds-3-{e}^{2}+2\,   \cos^2 f(1+e^2)-2\,e\cos  f\sin^2f }{\ds 2\,a\,e\,\sqrt {1-{e}^{2}}\sin  f  }} \\[3ex]
                                                                          {\frac {\ds\sqrt {1-{e}^{2}}\cos f   }{\ds a\,e}}\\[1.5ex]
                                                                          0 \\
                                                                        \end{array}
                                                                      \right]\;\;\,.
\end{equation}
Finally, we shall derive the gauge-generalised variation of the true anomaly. The relation
\begin{equation}\label{}
    r(1+e\cos f) =\frac{h^2}{GM}
\end{equation}
entails, upon differentiation with respect to $\vec r$,
\begin{equation}\label{}
    \frac{\p f}{\p \vec r} = -\frac{1}{e\sin f}\left[\frac{1}{r}\left(\frac{2h}{GM}\frac{\p h}{\p \vec r}-\frac{\p r}{\p \vec r}(1+e\cos f)\right)-\frac{\p e}{\p\vec r}\cos f\right]\;\;\;.
\end{equation}
Utilising the identity
\begin{equation}\label{}
    \frac{\p r}{\p \vec r} = \left[1,\,0,\,0\right]
\end{equation}
and plugging in equations (\ref{eq20}), (\ref{eq24}), we find that
\begin{equation}\label{dfdr}
    \left(\frac{\p f}{\p \vec r}\right)^T =  \frac{\sin f}{ep\cos f}\left[
                                                   \begin{array}{c}
                                                     -\sin  f (1+e\cos f)  \\[1.5ex]
                                                      (e^2\cos f+2e+\cos f) \\[1.5ex]
                                                     0 \\
                                                   \end{array}
                                                 \right]\;\;\,.
\end{equation}

\section*{Acknowledgments}

 The authors are grateful to Joseph O'Leary for a stimulating conversation, which moved the authors to return to an old draft and to turn it into a manuscript. One of the authors (ME) would like to thank Sergei M. Kopeikin for a useful consultation on the PN formalism.

\bibliographystyle{elsarticle-harv}
\bibliography{gr}

\end{document}